\documentclass[iop]{aastex}
\usepackage{color}
\usepackage{natbib}

\newcommand{\AKARI}{\textit{AKARI}}

\def\Teff{$T_{\rm{eff}}$}
\def\Tcr{$T_{\rm{cr}}$}
\def\Tcond{$T_{\rm{cond}}$}
\def\logg{log~\textit{g}}

\def\HtO{$\mathrm{H_{2}O}$}
\def\Ht{$\mathrm{H_{2}}$}
\def\CHf{$\mathrm{CH_4}$}
\def\COt{$\mathrm{CO_2}$}
\def\Nt{$\mathrm{N_2}$}
\def\NHt{$\mathrm{NH_3}$}
\def\TiO{$\mathrm{TiO}$}
\def\VO{$\mathrm{VO}$}
\def\AKARI{\textit{AKARI}}

\slugcomment{Not to appear in Nonlearned J., 45.}

\shorttitle{CO, CO$_2$ and CH$_4$ FUNDAMENTAL BANDS AND PHYSICAL PARAMETERS}
\shortauthors{Sorahana et al.}

\begin{document}

\title{{\textit{AKARI}} OBSERVATIONS OF BROWN DWARFS. III. CO, CO$_2$ and CH$_4$ FUNDAMENTAL BANDS AND PHYSICAL PARAMETERS}

\author{S. Sorahana$^{1,2}$ and I. Yamamura$^{2}$}
\affil{$^{1}$Department of Astronomy, Graduate School of Science, The University of Tokyo,
Bunkyo-ku, Tokyo 113-0033, Japan \linebreak
${^2}$Department of Space Astronomy and Astrophysics, Institute of Space and Astronautical Science (ISAS),\\ Japan Aerospace Exploration Agency (JAXA), 
Sagamihara, Kanagawa 252-5210, Japan}
\email{sorahana@ir.isas.jaxa.jp}

\begin{abstract}
We investigate variations in the strengths of three molecular bands, {\CHf} at 3.3~$\mu$m, CO at 4.6~$\mu$m and {\COt} at 4.2~$\mu$m, in 16 brown dwarf spectra obtained by {\AKARI}. Spectral features are examined along the sequence of source classes from L1 to T8. 
We find that the {\CHf} 3.3~$\mu$m band is present in the spectra of brown dwarfs later than L5, and the CO 4.6~$\mu$m band appears in all spectral types. 
The {\COt} absorption band at 4.2~$\mu$m is detected in late-L and T type dwarfs. 
To better understand brown dwarf atmospheres, we analyze the observed spectra using the Unified Cloudy Model (UCM).  
The physical parameters of the {\AKARI} sample, i.e., atmospheric effective temperature {\Teff}, surface gravity {\logg} and critical temperature {\Tcr}, are derived. 
We also model IRTF/SpeX and UKIRT/CGS4 spectra in addition to the {\AKARI} data in order to derive the most probable physical parameters. 
Correlations between the spectral type and the modeled parameters are examined. 
We confirm that the spectral type sequence of late-L dwarfs is not related to {\Teff}, but instead originates as a result of the effect of dust.
\end{abstract}


\section{Introduction}
Brown dwarfs are objects that are too light to sustain hydrogen fusion in their cores. 
Their effective temperatures are very low, ranging over 2200--600~K. 
They are classified into L and T spectral types. 
The discovery of the first genuine brown dwarf, Gl~229B, by \citet{Nakajima_1995}, triggered active study of these sources. With their intermediate masses and temperatures, brown dwarfs are expected to have the blended properties of stars and planets, bridging the gap between them. 
However, their properties (for example, their dusty atmospheres) make them unique enough to be classed separately, 
and it is not straightforward to understand their internal physical and chemical processes from our knowledge of stars and planets. 
Studies of brown dwarf atmospheres will lead us to a more comprehensive understanding of the nature of $``$atmospheres$"$ of various objects from stars to planets.

The photospheres of brown dwarfs are cool and dense (log $P_g$ $\sim$ 6.0 dyn cm$^{-2}$; $P_g$ is total gas pressure), and are thus dominated by molecules and dust. 
The chemistry of the photosphere and resultant molecular abundances govern the presence of spectral features.
Hydrogen is predominantly in the form of \Ht.
The dominant equilibrium form of carbon is CO and {\CHf}, 
oxygen is in {\HtO}, and nitrogen is in {\Nt} and {\NHt}.
Silicates, {\TiO} and {\VO} are found in objects with temperatures {\Teff} above 1600--2000~K \citep{Burrows_2001}. 
Neutral alkali metals are found at {\Teff} of $\sim$ 1000~K (\citealt{Fegley_1996}).
These values are derived by solving for thermo-chemical equilibrium. 

Condensation of dust under a thermo-chemical equilibrium photosphere was discussed as early as the 1960s \citep[e.g.][]{Lord_1965, Larimer_1967a,Larimer_1967b}. 
The major elements that construct dust grains are Fe, Mg, Si, O, Ca, Ti and Al. 
When temperature decreases below a certain threshold ($\sim$2000~K), condensation starts in the photosphere. 
For L dwarfs the dust exists in the upper photosphere and contributes to its spectral features. 
Dust in the photosphere contributes to the spectra directly by filling in the molecular absorption bands and by  extinction.
Dust also contributes indirectly by changing the thermal structure of the photosphere. 
On the other hand, for T dwarfs with lower {\Teff} dust disappears from the photosphere and does not play any role for the spectral features.    

Since almost all carbon atoms are in {\CHf} rather than CO in the photospheres of T dwarfs with {\Teff} less than about 1300~K under thermo-chemical equilibrium, 
it was our expectaction that the CO absorption band would not be present in the spectra of these coldest dwarfs. 
As we describe below, observations have shown us that this is not the case. 
Several observations from the ground have detected the CO absorption band at 4.6~$\mu$m in late-T dwarfs against theoretical expectation based on 
the local-thermodynamical equilibrium (LTE).
The band was observed in the T6 dwarf Gl~229B (\citealt{Oppenheimer_1998, Noll_1997}) 
and in the T8 dwarf Gl~570D (\citealt{Geballe_2009}). 
Another example of deviation from the thermal equilibrium chemistry was found in the nitrogen containing molecules observed by {\textit{Spitzer}} Space Telescope (SST).
The {\NHt} absorption band at 10.5~$\mu$m was much weaker than that expected from the atmospheric model (\citealt{Saumon_2006}). 
Although we do not know how common these phenomena are,  
these discrepancies between observed and model spectra have been critical problems in the study of brown dwarf atmospheres. 
To interpret the non-LTE abundances of these molecules, \citet{Griffith_1999} suggested that $``$vertical mixing$"$ plays a role,
in which CO molecules are dredged up from inner warm areas to outer cooler regions in the photosphere.  
However, the paucity of data to-date did not allow us to assess the relevance of this suggestion, and more spectroscopic data is required to investigate the above discrepancies.

Spectroscopic observations in the infrared regime are the most powerful tools to obtain physical and 
chemical information of brown dwarf photospheres, 
since brown dwarfs emit the majority of their radiative energy over this regime, and various molecular and dust features can be found therein. 
In the wavelength range of 2.5 to 5.0~$\mu$m, there are several prominent molecular absorption bands; 
{\CHf} $\nu_3$ fundamental band at 3.3~$\mu$m, {\COt} $\nu_3$ fundamental band at 4.2~$\mu$m, CO fundamental band at 4.6~$\mu$m and {\HtO} $\nu_1$ and $\nu_3$ absorption bands around 2.7~$\mu$m. 
The CO, {\CHf} and {\HtO} absorption bands are also present in the shorter wavelength range ($\leq$ 2.5~$\mu$m) 
and spectra of these bands have been used in previous studies of brown dwarf atmospheres. 
However, it is difficult to analyze these molecular bands independently because they are blended in the observed spectra. 
In addition, almost all absorption bands present at wavelengths shorter than 2.5~$\mu$m, for example CO at 2.3~$\mu$m and {\CHf} at 1.6 and 2.2~$\mu$m, are overtone bands, 
and are about 10--100 times weaker than the fundamental bands in 2.5--5.0~$\mu$m. 
These fundamental bands are mostly non-blended and suitable for the detailed analysis in the moderate resolution spectra. 
However, observations in this wavelength range from the ground are always challenging. 
Severe absorption due to the Earth's atmosphere and limited wavelength coverage make precise analysis difficult.

The Japanese infrared astronomical satellite {\AKARI} (\citealt{Murakami_2007}) was launched in February 2006. The
InfraRed Camera (IRC; \citealt{Onaka_2007}) on-board {\AKARI} is capable of yielding moderate-resolution ($R\sim120$)
spectra in this important wavelength range devoid of any degradation by telluric features. We
have conducted an observing program using the IRC to obtain continuous spectra of brown
dwarfs in 2.5--5.0~$\mu$m wavelengths aiming to carry out systematic studies of physical and
chemical processes in their atmospheres.
Continuous spectra of brown dwarfs in 2.5--5.0~$\mu$m were obtained by {\AKARI} for the first time, 
and provided new insight into the brown dwarf atmosphere. 

The initial results based on the {\AKARI} spectra of 6 brown dwarfs taken in the liquid-He cooled phase (Phase~2; see section \ref{akari}) are reported in \citet{Yamamura_2010} and \citet{Tsuji_2011}. 
\citet{Yamamura_2010} found that the observed CO band strength at 4.6~$\mu$m in late-L to late-T dwarfs are not consistent with predictions, and
attempted to explain the discrepancy of CO band strength in late-L to late-T dwarfs by vertical mixing effects. 
They argue that the CO band in late-T dwarfs could be reproduced by this effect, 
but earlier brown dwarfs between late-L and mid-T dwarfs are not. 
{\COt} absorption band at 4.2~$\mu$m in one L dwarf and two T dwarfs were also stronger than expected.
They find that the excess of {\COt} abundance can not be explained by vertical mixing either. 
Tsuji et al. (2011) suggested that a possible reason of 4.2~$\mu$m {\COt} absorption feature in the late-L and T type spectra 
is the higher than solar C and O elemental abundances used in the previous studies. 

In this paper, we summarize the observation and data reduction of {\AKARI} brown dwarf spectra in 2.5 -- 5.0~$\mu$m, 
and present the results of systematic analysis of 16 brown dwarf spectra covering a wide range of spectral types from L to T including those taken in the warm phase (Phase~3; see section \ref{akari}). 

\section{Observations and Data Reduction}

\subsection{{\AKARI}}
\label{akari}
{\AKARI} equipped with  an infrared telescope with an aperture of 68.5 cm. 
It was sensitive over the wavelength range from 1.7 to 180~$\mu$m with two scientific instruments; 
the Far-Infrared Surveyor (FIS; \citealt{Kawada_2007}) and the Infrared Camera (IRC). 
{\AKARI}'s primary mission was to carry out an all-sky survey in six bands, with a better sensitivity and spatial resolution than the previous survey by the IRAS mission (\citealt{Neugebauer_1984}). 
Thousands of pointed observations were also carried out. 
The liquid-He cool holding period of observations (Phase1, 2) lasted from 2006 May until 2007 August. 
After the boil-off of liquid-He, observations were continued with cryocooler only with the near-infrared camera of the IRC (Phase~3). 

\subsection{The InfraRed Camera (IRC) }
The InfraRed Camera (IRC) onboard {\AKARI} covers the wavelength range of 1.8--26.5~$\mu$m with three independent cameras operating simultaneously, 
namely the NIR (near-infrared), MIR-S (mid-infrared short), and MIR-L (mid-infrared long) channels.
Our observations were carried out in the AOT (Astronomical Observation Template) IRC04 for Phase~2 and IRCZ4 for Phase~3 with the observation parameter 
of $`$b;Np' (\citealt{Lorente_2008}). 
In this mode, the entire 2.5--5.0~$\mu$m wavelength range is covered with a grism with dispersion of 0.0097~$\mu$m/pixel 
or an effective spectral resolution of $R=\lambda/\mathnormal{\Delta} \lambda=$120 at 3.6~$\mu$m for point sources (\citealt{Ohyama_2007}). 
The source was placed in the 1$\times$1 arcmin$^{2}$ aperture, referred to as $`$Np',  
prepared for spectroscopy of point sources, preventing contamination of the spectra from nearby sources. 
A pointed observation by {\AKARI} allowed about 10 minutes exposure. 

\subsection{The Mission Program NIRLT}
{\AKARI} observation programs are classified into three categories, Large-Area Surveys (LS) organized by the project, Mission
Programs (MP) by the project members, and Open-Time programs (OT). 
We have conducted a Mission Program titled $``$Near-InfraRed spectroscopy of L and T dwarfs$"$ (NIRLT; P.I.  I. Yamamura) to obtain full NIR band spectra of brown dwarfs  using the IRC. 
The program aimed at constructing a set of legacy data for studies of the physical and chemical structure of brown dwarfs over a wide range of spectral types from L to T. 

Our target list included 40 objects selected by their expected fluxes 
(to be bright enough for the {\AKARI}/IRC instrument to provide high-quality spectra within a reasonable amount of exposure time) 
and their spectral types (to sample various types from M to T). 
Nine M dwarfs, seventeen L dwarfs, and fourteen T dwarfs were included in the target list. 

\begin{deluxetable}{lllrccc}
\tabletypesize{\scriptsize}
  \tablecaption{Observed L and T Objects in the NIRLT Program.\label{targetlist}}
\tablewidth{0pt}
\tablehead{
\colhead{Object Name} & \colhead{Object Name in This Paper} & \colhead{R.A. (J2000)} & \colhead{Decl.(J2000)} & \colhead{Sp. Type} & \colhead{Distance[pc]} & \colhead{Ref.}\\  
}
\startdata
2MASS~J14392837+1929150 &2MASS~J1439+1929& 14:39:28.40 &  +19:29:15.0 & L1 & 14.36&1,a \\
GD~165B &GD~165B& 14:24:39.09 & +09:17:10.4  & L3 & 31.69&2,a \\
Kelu--1 &Kelu--1& 13:05:40.20 & --25:41:06.0 & L3 & 18.74&2,a \\
2MASS~J00361617+1821104 &2MASS~J0036+1821& 00:36:15.90 &  +18:21:10.0 & L4& 8.76 &2,a\\
2MASS~J22244381--0158521 &2MASS~J2224--0158& 22:24:43.80 & --01:58:52.0 & L4.5 & 11.35&1,b \\
2MASS~J05395200--0059019 &2MASS~J0539--0059& 05:39:52.00  & --00:59:01.9  & L5 &  13.13&2,b\\
SDSS~J144600.60+002452.0 &SDSS~J1446+0024& 14:46:00.61   & +00:24:51.9      & L5 & 22.00&2,b \\
2MASS~J15074769--1627386 &2MASS~J1507--1627& 15:07:47.60 & --16:27:38.0 & L5 & 7.33&1,a \\
GJ~1001B &GJ~1001B& 00:04:36.40 & --40:44:03.0 & L5 & 9.55&1,a \\
2MASS~J08251968+2115521 &2MASS~J0825+2115& 08:25:19.60 &  +21:15:52.0 & L6 & 10.46& 2,b\\
2MASS~J17114573+2232044 &2MASS~J1711+2232& 17:11:45.73  & +22:32:04.4  & L6.5 &  30.34&1,b\\
2MASS~J16322911+1904407 &2MASS~J1632+1904& 16:32:29.10 &  +19:04:41.0 & L7.5 &  15.73&2,b\\
2MASS~J15232263+3014562 &2MASS~J1523+3014& 15:23:22.63      & +30:14:56.2      & L8 & 17.45&2,b \\
SDSS~J083008.12+482847.4 &SDSS~J0830+4828& 08:30:08.25      & +48:28:48.2      & L9 & 13.09&2,b \\
2MASS~J03105986+1648155 &2MASS~J0310+1648& 03:10:59.90 &  +16:48:16.0 & L9 & 25.00&2,c \\
2MASS~J03284265+2302051 &2MASS~J0328+2302& 03:28:42.60 &  +23:02:05.0 & L9.5& 30.32&2,b \\
SDSS~J042348.57--041403.5 &SDSS~J0423--0414& 04:23:48.60 & --04:14:04.0 & T0 & 15.24&2,b \\
SDSS~J125453.90--012247.4 &SDSS~J1254--0122& 12:54:53.90 & --01:22:47.0 & T2& 13.21& 2,b\\
SIMP J013656.5+093347.3 &SIMP J0136+0933& 01:36:56.60 &  +09:33:47.0 & T2.5 &  6.4&3,d\\
SDSS~J175032.96+175903.9 &SDSS~J1750+1759& 17:50:32.93     & +17:59:04.2  &T3.5 & 27.72&2,b \\
2MASS~J05591914--1404488 &2MASS~J0559--1404& 05:59:19.14      & --14:04:48.8  & T4.5 &  10.47&2,b\\
Gl~229B &Gl~229B& 06:10:34.70 & --21:51:49.0 & T6 & 5.8&2,a \\
2MASS~J15530228+1532369 &2MASS~J1553+1532& 15:53:02.20 &  +15:32:36.0 & T7 & 12.0&3,e \\
2MASS~J12171110--0311131 &2MASS~J1217--0311& 12:17:11.10   & --03:11:13.1 & T7.5 & 9.1&2,b \\
Gl~570D &Gl~570D& 14:57:15.00 & --21:21:51.0 & T8 & 5.9&2,f \\
2MASS~J04151954--0935066 &2MASS~J0415--0935& 04:15:19.54      & --09:35:06.6  & T8 &5.74& 3,b \\
$ \epsilon$ Ind Ba+Bb  &$ \epsilon$ Ind Ba+Bb& 22:04:10.52  & --56:46:57.7  &  T1+T6  &  3.62& 3,g  
\enddata
\tablecomments{Reference of spectral type (1) Kirkpatrick et al. (2000), 
(2) Geballe et al. (2002), 
(3) Burgasser et al. (2006).\\
Distances are estimated based on trigonometric parallaxes. 
The parallaxes are referred from (a) \citet{Dahn_2002}, (b) \citet{Vrba_2004}, (c) Stampf et al. (2010), (d) \citet{Artigau_2008}, 
(e) \citet{Jameson_2008}, 
(f) \citet{Burgasser_2000}, (g) \citet{King_2010}.\\
} 
\end{deluxetable}

\subsection{Observations}
\label{ob}
Thirty-five dwarfs (15 L, 11 T and 9 M) were observed, and  
thirty-three objects (14 L dwarfs, 10 T dwarfs and 9 M dwarfs) were detected. 
In addition, data of the L dwarf GJ~1001B were also obtained as part of the observation of its primary star, the M dwarf GJ~1001A. We list the 27 observed L and T dwarfs in Table~\ref{targetlist}. 
Table~\ref{datalist} summarizes all NIRLT observation records.
We observed 10 brown dwarfs in Phase~2, 
and 17 sources in Phase~3.


\begin{deluxetable}{lcccc}

\tablewidth{0pt}
\tablecaption{Summary of the Observations in the NIRLT program.\label{datalist}}
\tablehead{
\colhead{Object Name} & \colhead{Sp. Type} & \colhead{Date} & \colhead{ObsID:} & \colhead{Remarks} 
}
\startdata
2MASS~J14392+1929 & L1 & 2008.7.22 &1770009-001 & Wrong coordinate\\
2MASS~J14392+1929 & L1 & 2008.7.22 &1770009-002 & Wrong coordinate\\
2MASS~J14392+1929 & L1 & 2010.1.19 & 1771009-001& \\
2MASS~J14392+1929 & L1 & 2010.1.19 & 1771009-002& \\
GD~165B & L3 & 2007.7.24&1720074-001 & Data lost\\
GD~165B & L3 & 2008.7.22& 1770010-001& \\
GD~165B & L3 & 2008.7.22&  1770010-002& \\
GD~165B & L3 & 2010.1.20 & 1771010-001& Too faint\\
GD~165B & L3 & 2010.1.20 & 1771010-002& Too faint\\
GD~165B & L3 & 2010.1.20 &1771010-003 & Too faint\\
GD~165B & L3 & 2010.1.20 & 1771010-004& Too faint\\
Kelu--1 & L3  & 2008.7.16& 1770018-001&Wrong coordinates\\
Kelu--1 & L3  & 2008.7.16& 1770018-002&Wrong coordinates\\
2MASS~J0036+1821 & L4  & 2008.7.6&1770024-001 & \\
2MASS~J0036+1821 & L4  & 2008.7.6&1770024-002 & \\
2MASS~J0036+1821 & L4  & 2010.1.5& 1771024-001& \\
2MASS~J0036+1821 & L4  & 2010.1.5& 1771024-002& \\
2MASS~J2224--0158 &  L4.5  &2009.5.29 & 1770019-001& \\
2MASS~J2224--0158 &  L4.5  &2009.5.29 & 1770019-002& \\
SDSS~J0539--0059   & L5  & 2006.9.17&1720009-001 &   \\
SDSS~J0539--0059   & L5  & 2009.9.16&1770007-001 &   \\
SDSS~J1446+0024   & L5  & 2007.8.2& 1720072-001&   \\ 
2MASS~J1507--1627  & L5  & 2008.8.12& 1770020-001& \\
2MASS~J1507--1627  & L5  & 2009.2.7& 1770120-001& \\
GJ~1001B  & L5  & 2009.6.2& 1770036-001&  \\
GJ~1001B  & L5  & 2009.6.2& 1770036-002&  \\
2MASS~J0825+2115  & L6  & 2008.10.26&1770016-001 & \\
2MASS~J0825+2115  & L6  & 2009.4.23& 1770016-002& \\
2MASS~J0825+2115  & L6  & 2009.10.26& 1771016-001& \\
2MASS~J0825+2115  & L6  & 2009.10.26& 1771016-002& \\
2MASS~J1711+2232  & L6.5  & 2007.3.5& 1720001-001&  \\
2MASS~J1711+2232  & L6.5  &2008.9.5 & 1770001-001& Data lost \\
2MASS~J1711+2232  & L6.5  &2008.9.5 & 1770001-002& Data lost \\
2MASS~J1632+1904 & L7.5  & 2009.2.21&1770025-001 & Too faint\\
2MASS~J1632+1904 & L7.5  & 2009.2.21&1770025-002 & Too faint\\
2MASS~J1523+3014  & L8  & 2007.1.26& 1770002-001& \\
2MASS~J1523+3014  & L8  & 2008.7.30& 1770002-002& \\  
SDSS~J0830+4828  & L9  & 2006.10.20& 1720007-001&  \\ 
SDSS~J0830+4828  & L9  & 2006.10.21& 1720007-002&  \\ 
SDSS~J0830+4828  & L9  & 2009.4.17& 1770006-001&  \\ 
2MASS~J0310+1648  & L9 & 2008.8.12& 1770011-001& \\
2MASS~J0310+1648  & L9 & 2008.8.13& 1770011-002& \\
2MASS~J0328+2302  & L9.5  & 2009.8.19& 1770027-001&Too faint\\
2MASS~J0328+2302  & L9.5  & 2010.2.14& 1771027-001&Too faint\\
2MASS~J0328+2302  & L9.5  & 2010.2.14& 1771027-002&Too faint\\
SDSS~J0423--0414 & T0  & 2008.8.25& 1770015-001& \\
SDSS~J0423--0414 & T0  & 2008.8.25& 1770015-002& \\
SDSS~J1254--0122  & T2  & 2008.7.3& 1770012-001& Wrong coordinates\\
SDSS~J1254--0122  & T2  & 2008.7.4&1770012-002 & Wrong coordinates\\
SDSS~J1254--0122  & T2  & 2010.1.2& 1771012-001& \\
SDSS~J1254--0122  & T2  & 2010.1.3& 1771012-002& \\
SIMP J0136+0933  & T2.5  &2008.7.17 & 1770031-001& \\
SIMP J0136+0933  & T2.5  &2008.7.17 & 1770031-002& \\
SIMP J0136+0933  & T2.5  & 2010.1.16& 1771031-001& \\
SIMP J0136+0933  & T2.5  & 2010.1.16& 1771031-002& \\
SDSS~J1750+1759   &T3.5  & 2007.3.17& 1720050-001&Too faint \\
SDSS~J1750+1759   &T3.5  & 2007.3.17& 1720050-002& Too faint\\
2MASS~J0559--14044 & T4.5  & 2006.9.22& 1720006-001& \\
2MASS~J0559--14044 & T4.5  & 2006.9.22& 1720008-001& \\
2MASS~J0559--14044 & T4.5  & 2008.9.22& 1770005-001& \\
2MASS~J0559--14044 & T4.5  & 2008.9.22& 1770005-002& \\
Gl~229B  & T6  & 2008.9.25& 1770013-001& Contaminated\\
Gl~229B  & T6  & 2008.9.25& 1770013-002& Contaminated\\
2MASS~J1553+1532  & T7  & 2008.8.15& 1770022-001& \\
2MASS~J1553+1532  & T7  & 2008.8.15& 1770022-002& \\
2MASS~J1553+1532  & T7  & 2008.8.15& 1770022-003& \\
2MASS~J1553+1532  & T7  & 2009.2.9& 1770022-004& \\
2MASS~J1553+1532  & T7  & 2009.2.9& 1771022-001& Too faint\\
2MASS~J1553+1532  & T7  & 2009.2.9& 1771022-002& Too faint\\
2MASS~J1217--0311  & T7.5 & 2007.6.26& 1720068-001& Too faint\\ 
Gl~570D  & T8  & 2009.8.10& 1770023-001& \\
Gl~570D  & T8  & 2009.8.10& 1770023-002& \\
Gl~570D  & T8  & 2009.8.10& 1770023-003& \\
Gl~570D  & T8  &2009.8.11& 1770023-004& \\ 
2MASS~J0415--0935  & T8  & 2007.2.18& 1720005-001&Ghosting\\
2MASS~J0415--0935  & T8  & 2007.2.18& 1720005-002&Ghosting\\
2MASS~J0415--0935  & T8  & 2007.8.23& 5125080-001&DT\\
2MASS~J0415--0935  & T8  & 2007.8.24& 5125081-001&DT\\
2MASS~J0415--0935  & T8  & 2008.8.21& 1770004-001&Too faint\\
2MASS~J0415--0935  & T8  & 2008.8.21& 1770004-002&Too faint\\
2MASS~J0415--0935  & T8  & 2008.8.21& 1770004-003&Too faint\\
2MASS~J0415--0935  & T8  & 2008.8.21& 1770004-004&Too faint\\
$\epsilon$ Ind Ba+Bb  &  T1+T6   & 2006.11.2& 1720003-001&\\    
$\epsilon$ Ind Ba+Bb  &  T1+T6   & 2006.11.2& 1720004-001&\\    
$\epsilon$ Ind Ba+Bb  &  T1+T6   & 2008.11.1& 1770003-001&\\    
$\epsilon$ Ind Ba+Bb  &  T1+T6   & 2008.11.1& 1770003-002&\\   
\enddata
\tablecomments{\\
Wrong coordinates: Observed with wrong coordinates\\
Data lost: Data downlink failed due to troubles in the ground system\\
Too faint: The objects was too faint\\
Contaminated: Not able to extract source spectrum due to heavy contamination from the nearby bright star\\
Ghosting: The data were not obtained due to instrumental ghosting.\\
DT: Observed as part of Director's Time}
\end{deluxetable}

\subsection{Data Reduction}
The standard software toolkit $IRC$$\_$$SPEC$$\_$$TOOLKIT$ (\citealt{Ohyama_2007}) was used for data reduction. 
We used the toolkit version 20110301, released in 2011 March. 
Wavelength and flux calibrations were all done automatically in the toolkit. 
Spectral data are extracted from two dimensional spectral images.  
The two axes of the images correspond to spatial and wavelength directions, respectively. 
Along the spatial direction the signal is extended by the point spread function (PSF), 
and along the wavelength direction the signal extension is determined by the spectral resolution and the PSF. 
The typical wavelength calibration error is 0.5 pixel of the detector or $\sim$0.005$\mu$m (\citealt{Ohyama_2007}), 
but could be larger in some cases (see Table~\ref{xshift}). 
We applied small corrections (0.01--0.03$\mu$m) to the data of several sources by comparing the position of the {\CHf} Q-branch feature with other objects. 
The overall flux calibration error is 10 $\%$ in the middle of the wavelength range, and 20 $\%$ at the short/long wavelength edges. 

We carried out the following three additional processing steps in order to improve the final data quality; 
(1) derivation of appropriate sky background, 
(2) stacking of multiple observations, 
and (3) correction of contaminating light from nearby objects.  

\subsubsection{Derivation of Probable Sky Background}
We modified the $IRC\_ SPEC\_ TOOLKIT$ program to improve the sky subtraction. 
The original program subtracts the sky derived from same pixel width with the on source signal. 
Since the Phase~3 data are noisier than those in the Phase~2, the sky level derived from only a few pixels (3--5 pixels) are not sufficiently flat. 
The revised program derives the sky level using a larger area ($\sim$10 pixels).

\begin{deluxetable}{lcrccc}
\tabletypesize{\scriptsize}
\tablecaption{Summary of data reduction.  \label{xshift}}
\tablewidth{0pt}
\tablehead{
\colhead{Object Name} & \colhead{Sp. Type} &  \colhead{x shift[$\mu$m]} & \colhead{number of pointings}&\colhead{total number of flames} &\colhead{remarks}
}
\startdata
2MASS~J1439+1929 & L1 & 0.00 & 2&14& \\
GD~165B & L3 & 0.00&--&--&Too faint  \\
Kelu--1 & L3 & -- & -- &--& Not detected \\
2MASS~J0036+1821 & L4 & 0.015 & 2&15 &\\
2MASS~J2224--0158 & L4.5 & --0.010 & 2&16& \\
SDSS~J0539--0059 & L5 & 0.00 & 1&9&   \\
SDSS~J1446+0024 & L5 & 0.00 & 1&9  & \\ 
2MASS~J1507--1627 & L5 & --0.020 & 1&7 &\\
GJ~1001B & L5 &  0.00 & 1&9 &\\
2MASS~J0825+2115 & L6 & 0.00 & 4&31 &\\
2MASS~J1711+2232 & L6.5 &  0.00 & 1&9& Too faint \\
2MASS~J1632+1904 & L7.5 & 0.00 & 2&15&\\
2MASS~J1523+3014 & L8 & --0.030 & 1&9 &\\
SDSS~J0830+4828 & L9 & 0.00 & 1&9  &\\ 
2MASS~J0310+1648 & L9 & 0.00 & 2&12 &\\
2MASS~J0328+2302 & L9.5 & --& --&--& Too faint\\
SDSS~J0423--0414 & T0 & 0.00 & 1&9& Binary \\
SDSS~J1254--0122 & T2 & 0.00 & 1&9& \\
SIMP J0136+0933 & T2.5 & 0.010 & 4&31& \\
SDSS~J1750+1759&T3.5 & --&--&--& Too faint   \\
2MASS~J0559--14044 & T4.5 & 0.00 & 1&9&  \\
Gl~229B & T6 & -- & -- &--& Contamination\\
2MASS~J1553+1532 & T7 & --0.010 & 4&24& \\
2MASS~J1217--0311 & T7.5 & --&--&--& Too faint \\
Gl~570D & T8 & 0.00 & 4&28& \\
2MASS~J0415--0935 & T8 & --0.020 & 1&9 & \\  
ƒÃ Ind Ba+Bb  &  T1+T6  &  0.00&1&9 & Binary  
\enddata
\end{deluxetable}

\subsubsection{Stacking of Multiple Observations}
Since brown dwarfs are generally very faint (a few $\mathrm{mJy}$ $< F_{\nu} <$ 25 $\mathrm{mJy}$), 
one pointed observation is not always sufficient to obtain good quality data. 
This is especially true for the Phase~3 observations, where the noise level is about a factor of 2.5 higher than that of Phase~2. Six to eight spectral frames were taken per pointing (\citealt{Lorente_2008}). 
The toolkit stacks all available exposure frames within a pointing. 
We observed at least twice per object, unless the observations failed for some reason. 
The toolkit does not stack the frames over multiple pointings. 
For this, we used additional custom-made programs; $IRC\_ SPEC\_ TOOLKIT\_ wSTACMULTI$ version 20100918 (Shimonishi priv. comm.). 
The fourth and fifth columns in Table~\ref{xshift} show the number of pointings and the total number of frames used in the data reduction, respectively.
The stacked data are better than single pointing data. 
When an object was observed both in Phase~2 and Phase~3, we used only Phase~2 data.

\subsubsection{Subtraction of Signal from a Nearby Object}
This process was partially applied to GJ~1001B. 
GJ~1001 is a low-mass binary system, with GJ~1001B being the companion of the M dwarf, GJ~1001A. 
The difference in the magnitude is about 3 mag at $L'$ band, e.g. GJ~1001A is about 16 times brighter than GJ~1001B. 
Since the separation between GJ~1001B and GJ~1001A is only 13 arcsec (11 pix on the detector), 
the spectrum of GJ~1001B was contaminated by a shoulder of intense signal from GJ~1001A as the PSF of the IRC/NIR channel has a FWHM $\sim$ 3.2 pix. 
To measure the signal of GJ~1001B accurately, we took account of the extended signal of GJ~1001A. 
We assumed that the spectral image of GJ~1001A was symmetric in the spatial direction with respect to the flux peak pixels, 
and measured the $``$background$"$ level at the same distance from GJ~1001A to GJ~1001B on the opposite side. 
Then we subtract that value from the spectra of GJ~1001B instead of the background determined by the normal processing.

\subsection{Validation of Absolute Flux Calibration}
\label{aflux}
Among the observed 25 brown dwarfs (15 L dwarfs and 10 T dwarfs) by {\AKARI},  
16 sources (11 L dwarfs and 5 T dwarfs) present high quality spectra  
whose averaged signal to noise ratio (S/N) is higher than about 3.0. 
The corresponding flux level is about 1 mJy for the Phase~2 data and 2.5 mJy for the Phase~3 data.
The highest S/N and mean S/N are about 18 and 8, respectively. 
Four known binary brown dwarfs, $\epsilon$ Ind Ba+Bb (T1+T6), SDSS~J0423--3014 (T0), 2MASS~J0310+1648 (L9)  and 2MASS~J1553+1532 (T7), are excluded. 
The dataset is summarized in Table~\ref{maggol_l}.
Six objects taken in Phase~2 and 10 objects in Phase~3 are used for the analysis in this paper. 

\begin{figure}
\epsscale{.80} 
\begin{center}
\plotone{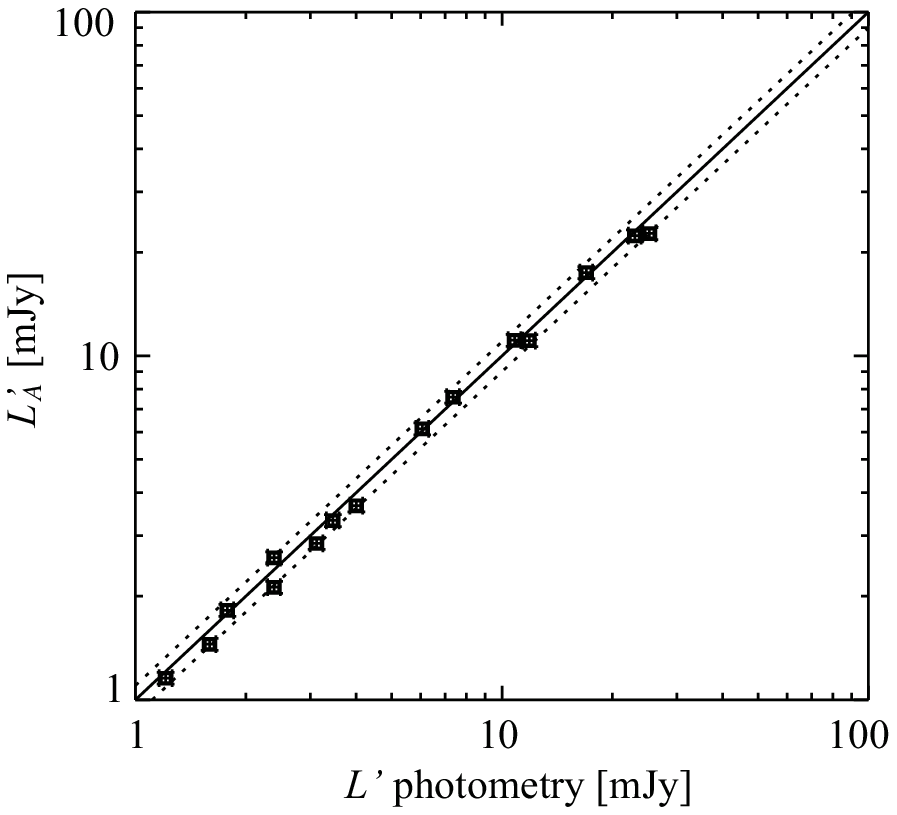}
\end{center}
\caption{The comparison of the $L'$ flux from {\AKARI} spectra and $L'$ photometry (Leggett et al. 2002a,b; Golimowski et al. 2004). The past $L'$ photometric values are converted to $F_{\nu}$ [mJy]. 
The border of $\pm 10 \%$ is denoted by the dotted lines.
We see that the $L'$ fluxes from {\AKARI} data are consistent with past $L'$ photometry to within 10 $\%$.
}\label{L}
\end{figure}

We checked {\AKARI}'s absolute flux calibration by comparing the $L'$ band fluxes with past photometry from \citet{Leggett_2002a,Leggett_2002b} and \citet{Golimowski_2004} (Table~\ref{maggol_l}). 
We derive the $L'$ fluxes from {\AKARI} spectra by applying the Mauna Kea Observatory (MKO) $L'$ filter 
used by \citet{Leggett_2002a,Leggett_2002b} and \citet{Golimowski_2004}. 
The 50 $\%$ cut-off wavelength of the filter is 3.43~$\mu$m and 4.11~$\mu$m. 
No previous $L'$ photometry data is available for SIMP J0136+0933. 
Figure~\ref{L} shows the comparison of the $L'$ flux from {\AKARI} (hereafter $L'_{A}$) and past $L'$ photometry values. 
We see that the $L'_{A}$ are consistent with the past $L'$ photometry to within 10 $\%$. 

\begin{deluxetable}{lccccccc}
\tabletypesize{\scriptsize}
\tablecaption{Infrared magnitudes and colors of the analyzed brown dwarfs.
 \label{maggol_l}}
\tablewidth{0pt}
\tablehead{
\colhead{Object Name} & \colhead{Sp. Type} &  \colhead{$L'$} &  \colhead{$L'_A$} &  \colhead{[3.3]--$L'_A$} &  \colhead{$J$--$L'_A$}&  \colhead{$J$--$K$}&  \colhead{$J_{\mathrm{p}}$}
}
\startdata
2MASS~J1439+1929& L1&10.80$^1$&10.87(0.003)& 0.47&1.79&1.20&12.66$^1$\\
2MASS~J0036+1821 & L4 &  10.08$^1$&10.11(0.002)& 0.46&2.20&1.36&12.31$^1$\\
2MASS~J2224--0158 & L4.5 & 10.90$^3$&10.87(0.002)& 0.45&3.02&1.97&13.89$^5$ \\
GJ~1001B& L5 & 10.41$^2$&10.38(0.004)& 0.47&2.68&1.68&13.06$^1$ \\
SDSS~J1446+0024& L5 & 12.54$^3$&12.67(0.004)& 0.44&2.89&1.81&15.56$^1$ \\ 
SDSS~J0539--0059 & L5  & 11.32$^1$&11.29(0.001)& 0.57&2.56&1.37&13.85$^1$ \\
2MASS~J1507--1627& L5 & 9.98$^1$&10.10(0.002)& 0.58&2.60&1.45&12.70$^1$\\
2MASS~J0825+2115& L6  &11.53$^1$&11.52(0.004)& 0.80&3.37&1.94&14.89$^1$\\
2MASS~J1632+1904 & L7.5  & 12.54$^3$&12.45(0.009)& 0.96&3.32&1.81&15.77$^1$\\
2MASS~J1523+3014& L8  &12.86$^1$&12.84(0.004)& 0.90&3.11&1.82&15.95$^1$ \\
SDSS~J0830+4828 & L9  &  11.98$^1$&12.08(0.003)& 1.11&3.14&1.61&15.22$^1$\\ 
SDSS~J1254--0122& T2 & 12.25$^3$&12.35(0.010)& 1.48&2.31&0.93&14.66$^1$\\
SIMP J0136+0933 & T2.5 &  N/A$^4$&10.88(0.003)& 1.58&2.23&0.71&13.46$^6$\\
2MASS~J0559--1404 & T4.5 &  12.14$^1$&12.18(0.003)& 2.55&1.39&--0.06&13.57$^1$\\
Gl~570D& T8  & 12.98$^3$&13.08(0.016)& 2.02&1.74&--0.58&14.82$^1$\\
2MASS~J0415--0935 & T8  &13.28$^3$&13.33(0.007)& 3.56&1.99&--0.49&15.32$^5$
\enddata
\tablecomments{The error in $L'$ and $J_{\mathrm{p}}$ is typically 5~\%. 
SIMP J0136+0933 is calibrated with 2MASS~photometric data.
$L'$ and $J_{\mathrm{p}}$ magnitudes are from $^1${\citet{Leggett_2002a}}, 
$^2${\citet{Leggett_2002b}},
$^3${\citet{Golimowski_2004}},
$^4${No data},
$^5${\citet{Knapp_2004}},
$^6${\citet{Cutri_2003}}}.
\end{deluxetable}

\section{The 2.5 -- 5.0~$\mu$m Spectral Dataset of Brown Dwarfs}
\label{template}
\begin{figure}
\epsscale{.70}
\begin{center}
\plotone{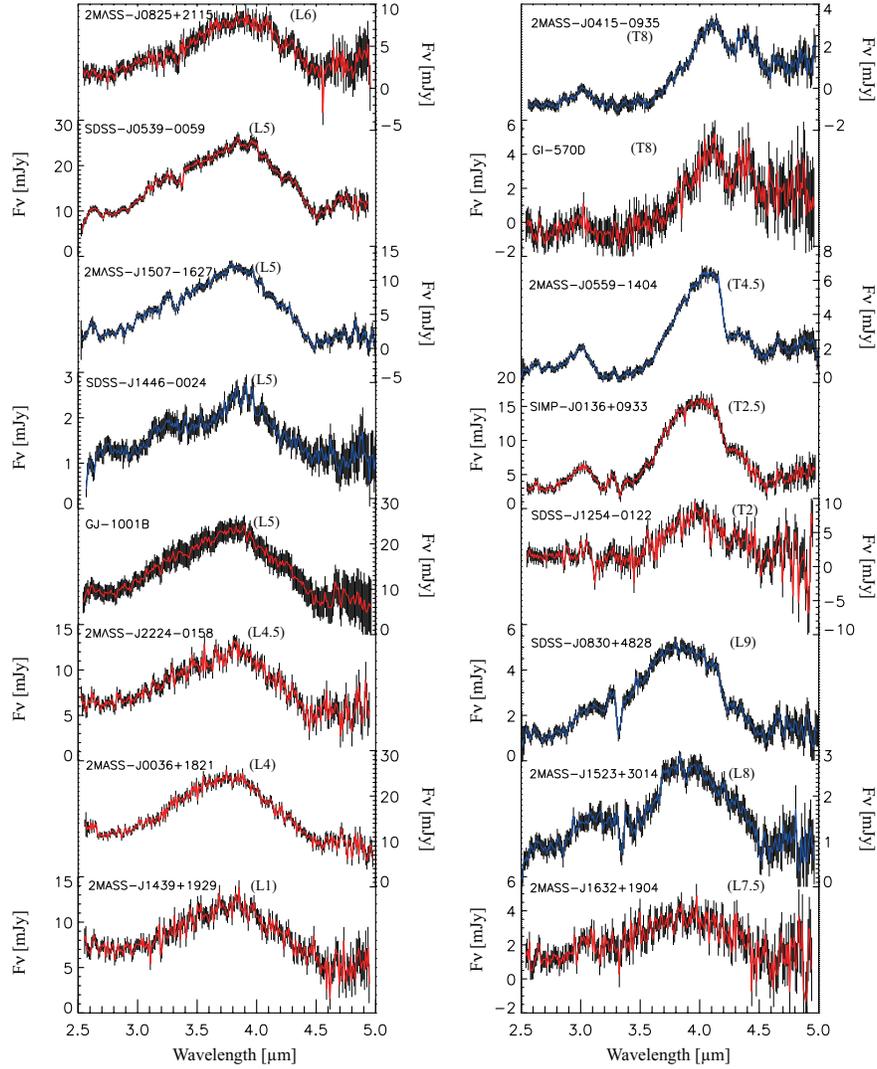}
\end{center}
\caption{{\AKARI} spectra of 11 L-dwarfs and 5 T-dwarfs. 
Data taken in Phase~2 are shown as the blue line and these in Phase~3 are drawn in red. The Phase~2 data are generally of better quality than the Phase~3 data. 
The CO 4.6~$\mu$m band appears in the spectra of all spectral types.
The {\CHf} 3.3~$\mu$m fundamental band appears in the spectra later than L5. 
The band is seen in SDSS~J0539--0059 and 2MASS~J1507--1627, 
but not in the other two L5 sources,  GJ~1001B and SDSS~J1446+0024.
The {\COt} absorption band at 4.2~$\mu$m presents in the spectra of late-L and T type dwarfs.
The band  appear clearly in the spectra of 2MASS~J0825+2115 (L6) and SDSS~J0830+4828 (L9), but not in the spectrum of  2MASS~J1523+3014(L8).
  } \label{allspectra}
  \end{figure}

Figure~\ref{allspectra} shows the spectra of the brown dwarfs in the sequence of their spectral types from L (left bottom) to T (right top). 

The following molecular absorption bands are clearly recognized in the {\AKARI} data; {\HtO} (broad absorption bands around 2.7~$\mu$m and at longer than 4.0~$\mu$m), CO (4.6~$\mu$m), {\COt} (4.2~$\mu$m), and {\CHf} (3.3~$\mu$m). 
Figure~\ref{allspectra} shows identification of these bands in the {\AKARI} spectra. 
The {\AKARI} spectra of L-type dwarfs are gently peaking at 3.8~$\mu$m, 
and are rather smooth and featureless with the current resolution throughout 2.5--5.0~$\mu$m except for the positions of H$_{2}$O and CO.  
{\CHf} at 3.3~$\mu$m exists in the dwarfs later than L5. 
On the other hand, the spectra of T-type dwarfs exhibit deep molecular absorption features further to the broad H$_{2}$O bands and CO band.

\subsection{CO Absorption Band at 4.6~$\mu$m}
\label{coabsorption}
Under the assumption of LTE, CO is expected to disappear from the spectra of the late-T type dwarfs 
because carbon resides mostly in {\CHf} rather than in CO in a very cool and high-density (T$\sim$1000~K, log $P_g$ $\sim$ 6.0 dyn cm$^{-2}$) environment (Tsuji 1964). 
Only few brown dwarf spectra have been obtained in the $M$ band wavelength range (\citealt{Noll_1997, Oppenheimer_1998, Geballe_2009}). 
They all showed the detections of the CO absorption band in the spectra of mid- to late-T dwarfs. 
These results raised a very important problem regarding the physics and chemistry of brown dwarf atmospheres. 
However, due to disturbance caused by the Earth's atmosphere the quality of these  data was not sufficient and the wavelength coverage was limited. 
Thus detailed analysis of these bands was not easy. 

We have obtained much better spectra covering broader range including the CO band with {\AKARI}. 
The {\AKARI} data show that CO appears in the all observed brown dwarf spectra from early-L to late-T type. 
We confirm that the spectra of our T-type brown dwarfs clearly exhibit CO absorption band, 
supporting the previous ground-based studies. 
It is now clear that the presence of CO in the late-T dwarfs is a common characteristic. 
It has been argued that CO in the photosphere of late-T dwarfs is maintained by vertical mixing (\citealt{Griffith_1999,Saumon_2000,Yamamura_2010}). 
The vertical mixing transfers CO molecules from the inner regions, where CO is still abundant, to outer cooler regions in the photosphere. 
However this mechanism does not fully explain the observed strength of the CO band, 
especially for the late-L to early-T dwarfs (Yamamura et al. 2010). 
In this paper we do not discuss the strength of CO band any more. 

\subsection{{\COt} Absorption Band at 4.2~$\mu$m}
\label{cotres}
{\AKARI} detected {\COt} absorption band at 4.2~$\mu$m in the spectra of brown dwarfs. 
The band is recognized in all T dwarfs and some late-L dwarfs. 
We see the band in the spectra of 2MASS~J0825+2115 (L6) and SDSS~J0830+4828 (L9) clearly, 
but not in the spectra of 2MASS~J1523+3014(L8).  

\begin{figure}
  \begin{center}
   \plotone{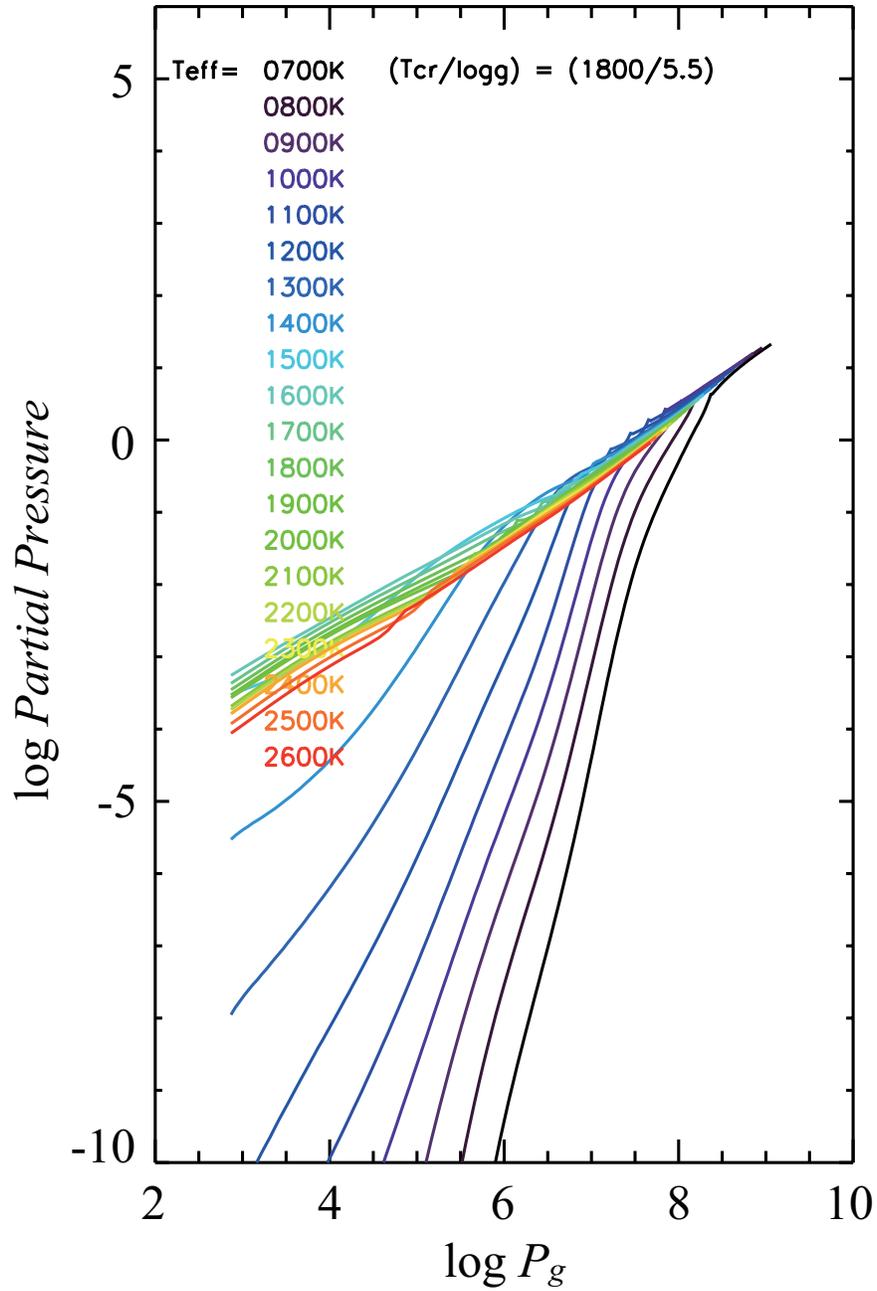}
  \end{center}
  \caption{ Partial pressures of {\COt} molecule for different {\Teff} are plotted against total gas pressure.  
{\COt} partial pressure in photosphere has a peak at {\Teff}=1600~K. 
  } \label{pp}
  \end{figure}

We investigate which spectral types of objects show the {\COt} band. 
Figure~\ref{pp}(a) shows the variation of partial pressure of {\COt} for different effective temperatures ({\Teff}) in the models under the LTE assumption. 
We see that the partial pressure of {\COt} increases with decreasing {\Teff} from 2600 to 1600~K, 
then changes to a decreasing trend with decreasing {\Teff} from 1600 to 700~K. 
This indicates that {\COt} is the most abundant in a photosphere with {\Teff} $\sim$1600~K.  
Thus the {\COt} absorption band should appear in spectra from L6 to early T dwarfs. 
However, {\COt} 4.2~$\mu$m absorption band does not appear in some observed spectra of late-L to early-T. 
This analysis tells us that the behavior of {\COt} absorption at late-L to early-T dwarfs is very complicated. 
It may reflect the difference in the elemental abundances (\citealt{Tsuji_2011}, Sorahana et al. in prep.). 
However, the appearance of {\COt} absorption band for late-T dwarfs can not be explained.

\subsection{{\CHf} Absorption Band at 3.3~$\mu$m}
\label{chfab}
It is known that the {\CHf} $\nu_3$ absorption band at 3.3~$\mu$m already appears in the spectra of a mid-L dwarf (Noll et al. 2000), 
but the {\CHf} $\nu_2+\nu_3$ absorption band at 2.2~$\mu$m which is used for classification of T-type dwarfs, 
was not detected in the spectra of the same L dwarf (Nakajima et al. 2004). 
The {\AKARI} data including 3.3~$\mu$m region should enhance our understanding of the CH$_{4}$ molecule 
in the photospheres of L dwarfs. 
We find that the {\CHf} 3.3~$\mu$m fundamental band appears in the spectra of brown dwarfs later than L5. 
Interestingly, we see the band in only two sources out of four L5 dwarfs in our {\AKARI} sample. 
The band is seen in SDSS~J0539--0059 and 2MASS~J1507--1627, 
but not in the other two L5 sources, SDSS~J1446+0024 and GJ~1001B (Figure~\ref{allspectra}). 

\subsubsection{Equivalent Width of the CH$_4$ Absorption Band}
\begin{figure}
\begin{center}
\plotone{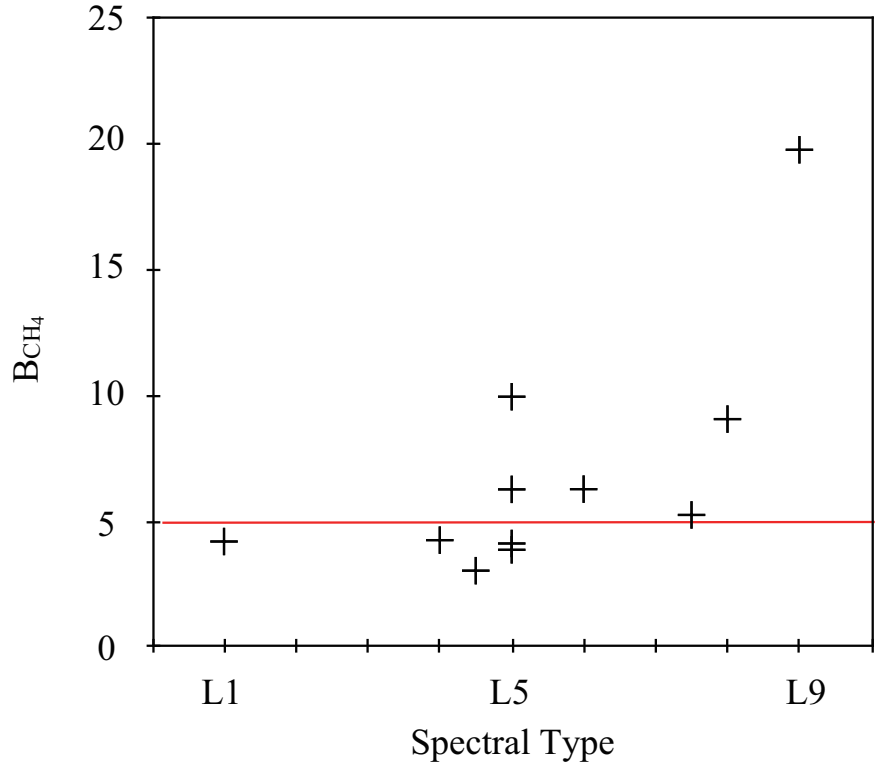}
\end{center}
\caption{ {\CHf} index $B_{\mathrm{CH_4}}$ measured on the {\AKARI} spectra is plotted versus spectral type. $B_{\mathrm{CH_4}}$ is ratio of the equivalent width of {\CHf} Q-branch to its uncertainty from the standard deviation of the data points in the nearby off-band wavelengths. It is found that the onset of the {\CHf}  feature is L5-type dwarfs. The red line indicates the threshold of {\CHf} detection. 
} 
\label{methan}
\end{figure}

We examine the appearance of the CH$_{4}$ band quantitatively in the {\AKARI} spectra from L1 to L9 dwarfs to confirm this result in detail. 
In L dwarfs the {\CHf} band is still weak and only the Q-branch feature is prominent. 
We derive the equivalent width of the 3.3~$\mu$m CH$_{4}$ Q-branch feature in each spectrum, 
and calculated the ratio between the equivalent width and its uncertainty derived from the standard deviation of the data in the nearby off-feature wavelengths. 
We evaluate $``${\CHf} index$"$ $B_{\mathrm{CH_4}}$ as below. 

\begin{equation}
\label{eweq}
B_{\mathrm{CH_4}}=\frac{EW}{\left(\frac{\sigma}{F_{\mathrm{center}}}\times d\lambda \times \sqrt{N-1} \right)}, 
\end{equation}
where EW is the equivalent width, $\sigma$ is the standard deviation in the off-band wavelengths, 
$F_{\mathrm{center}}$ is the estimated flux at the wavelength of the band center derived by linear interpolation from the off-feature region fluxes, 
$d\lambda$ is the wavelength grid interval and $N$ ($\sim$ 20) is the number of data points within the defined region.
We show the results in Table~\ref{ew} and Figure~\ref{methan}. 
We regard the detection to be significant when $B_{\mathrm{CH_4}}$ is larger than 5. 
This threshold is chosen because of the following reasons; 
the $B_{\mathrm{CH_4}}$ of the L7.5 dwarfs where the {\CHf} band is confirmed by eye is 5.25. For L1 and L4 dwarfs where the band is not confirmed are 4.19 and 4.24, respectively. 
We find that the {\CHf} 3.3~$\mu$m Q-branch feature starts appearing at L5 type, 
and the band appears in only two of four L5 dwarfs, SDSS~J0539--0059 and 2MASS~J1507--1627.  
The detection of the {\CHf} absorption band in the spectra of 2MASS~J1507--1627 is consistent with the past result reported by \citet{Noll_2000}.

\begin{deluxetable}{lcc}
\tabletypesize{\scriptsize}
\tablecaption{The {\CHf} index for L dwarfs. \label{ew}}
\tablewidth{0pt}
\tablehead{
\colhead{Object Name} & \colhead{Sp. Type} & \colhead{$B_{\mathrm{CH_4}}$} 
}
\startdata
2MASS~J1439+1929&L1&4.12 \\
2MASS~J0036+1821 &L4&4.24 \\
2MASS~J2224--0158 &L4.5& 3.03 \\
GJ~1001B&L5& 3.86 \\
SDSS~J1446+0024&L5& 4.11\\ 
SDSS~J0539--00590 &L5&9.95  \\
2MASS~J1507--1627&L5& 6.26\\
2MASS~J0825+2115&L6& 6.27\\
2MASS~J1632+1904 &L7.5&5.25\\
2MASS~J1523+3014&L8& 9.06 \\
SDSS~J0830+4828 &L9&19.78 
\enddata
\tablecomments{
We define the detection threshold as 5.0.}
\end{deluxetable}

\subsubsection{Color index [3.3]--$L'_A$}
\begin{figure}
\begin{center}
   \plotone{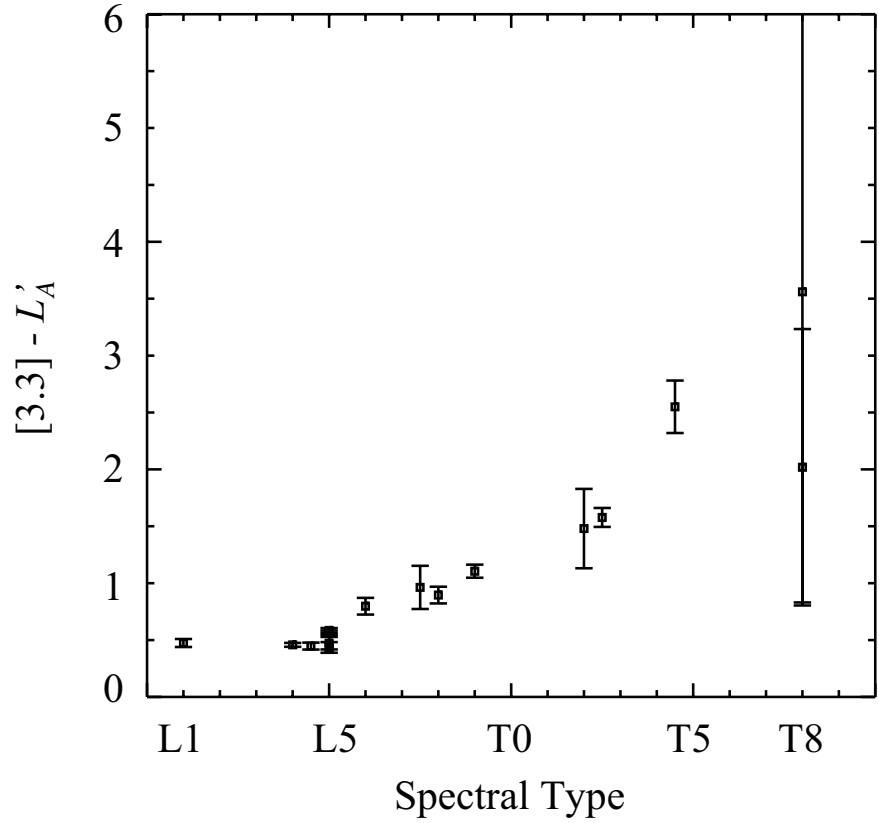}
\end{center}
\caption{[3.3]--$L'_A$ color as a function of spectral class. 
Whereas the colors of early type dwarfs do not change, the color of later than L5 dwarfs tend increasing. 
This indicates that the {\CHf} absorption band at 3.3~$\mu$m becomes stronger toward late spectral types. } 
\label{ln33}
\end{figure}

$B_{\mathrm{CH_4}}$ used in the previous section is only applicable for L-type dwarfs. 
The {\CHf} absorption band in T-type spectra becomes broader and deeper, and P- and R- branches are not negligible any more. 
In order to follow the variation of the {\CHf} 3.3~$\mu$m absorption in the spectra of brown dwarfs including T-type, 
we define a photometry band [3.3], 
which is measured by simply averaging the flux between  3.27 and 3.36~$\mu$m. 
[3.3]--$L'_A$ are listed in Table~\ref{maggol_l},  
and  plotted against the spectral types in Figure~\ref{ln33}. 
The large error of the color for late-T type sources is caused by lack of valid data points around 3.3~$\mu$m in the {\AKARI} spectra due to heavy {\CHf} absorption. 
[3.3]--$L'_A$ colors of early-L dwarfs do not change because {\CHf} absorption band at 3.3 $\mu$m does not appear in these cases. 
The [3.3]--$L'_A$ increases monotonically along L to T sequence, 
indicating that the {\CHf} $\nu_3$ absorption develops simply toward the later spectral types. 

\subsubsection{$J - K$ color}
Next, the effects of dust on the observed {\CHf} band strength are evaluated.
The $J$ band flux is the most sensitive to dust extinction. 
Thus the $J-K$ color would give information of the conditions of dust in the L dwarf photospheres. 
$K$ photometric values are obtained from \citet{Leggett_2000, Leggett_2001, Leggett_2002a, Leggett_2002b} and \citet{Knapp_2004}. 
$J - K$ colors are shown in Table~\ref{maggol_l}. 
We see a trend of redder colors from L1 to L6. Thereafter, $J-K$ colors become bluer later than L6. 
The red color of early- to mid-L dwarfs is thought be due to increasing dust extinction at $J$ band. 
We find that the $J-K$ colors of two L5 dwarfs showing the {\CHf} $\nu_3$ absorption band, 2MASS~J1507--1627 and SDSS~J0539--0059, are bluer than that of other L5 objects without the band, SDSS~J1446+0024 and GJ~1001B. 
This indicates that the difference in L5 dwarfs with or without {\CHf} $\nu_3$ absorption band is caused by dust reddening. 

\begin{figure}
\begin{center}
      \plotone{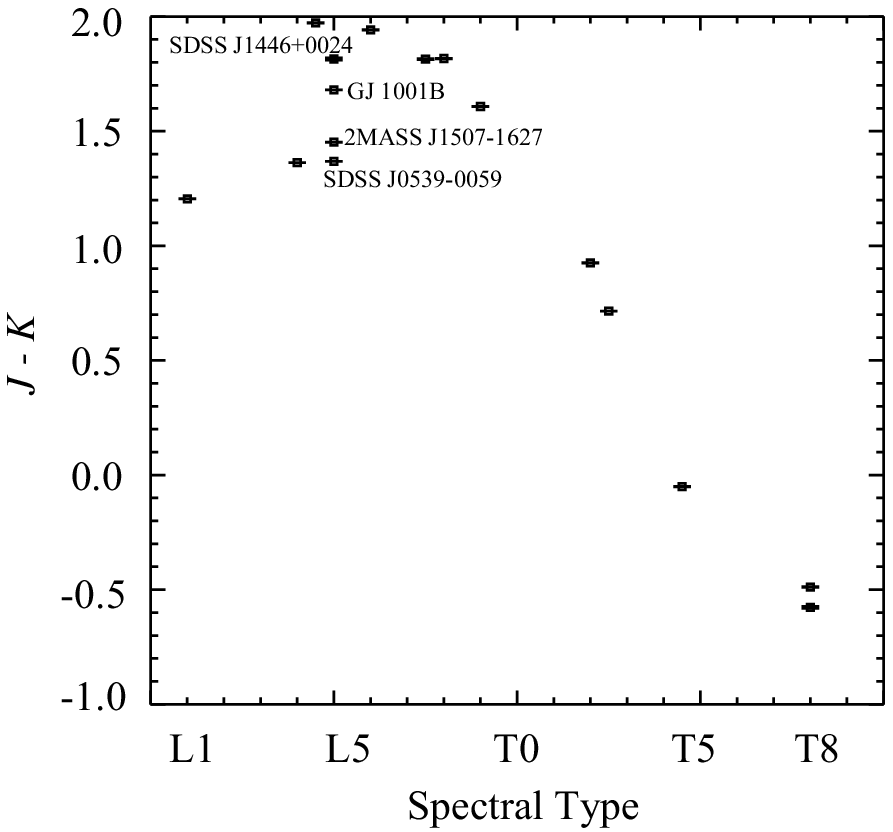}
\end{center}
\caption{$J-K$ color for objects observed by {\AKARI}. The color becomes redder from early- L to late-L dwarfs. 
The color from late-L to late-T becomes bluer. 
} 
\label{jkcolor}
\end{figure}

\section{Interpretation with the UCM}

\subsection{The Unified Cloudy Model}
\label{ucm}
To understand the atmospheres of brown dwarfs better, 
we analyze the {\AKARI} spectra using the Unified Cloudy Model (UCM, \citealt{Tsuji_2002,Tsuji_2005}). 
The atmospheres of cool stars are dominated by molecules. 
The UCM calculates molecular abundance by assuming LTE. 
Generally, stellar spectra can be interpreted in terms of effective temperature $T_{\rm{eff}}$, surface gravity log \textit{g}, chemical composition, and micro-turbulent velocity. 
Dust is an essential component in the atmospheres of brown dwarfs.
We assume metallic iron, enstatite (MgSiO$_3$) and corundum (Al$_2$O$_3$) in the UCM as the dust species.
Under the LTE dust forms at a layer where temperature drops down to the condensation temperature, $T_{\rm{cond}}$. 
Although we do not know the exact physics behind the behavior of dust layers, 
comparison with observations tells that it is difficult to explain the spectra with only these four basic parameters. 
The UCM assumes that the dust disappears at somewhat lower temperatures given as an additional parameter, namely the critical temperature $T_{\rm{cr}}$. 
Thus the dust would exist only in the layer with $T_{\rm{cr}}  < T <  T_{\rm{cond}}$. 
$T_{\rm{cr}}$ is not predictable by any physical theory at present and is required to be determined from observations empirically. 
The UCM apply the CH$_{4}$ line list by \citet{Freedman_2008}, 
which is based on the Spherical Top Data System (STDS) model of \citet{Wenger_1998} and 
believed to be the best one currently available for this band.
Other line lists are CO  (\citealt{Guelachvili_1983}, \citealt{Chackerian_1983}), 
{\COt} (HITEMP database; \citealt{Rothman_1997})
and {\HtO} (\citealt{Partridge_1997}).
See \citet{Tsuji_2002, Tsuji_2005} for details of the model. 
We assume that the objects have solar metallicity (here we take the values provided by \citealt{Allende_2002}) and that the micro turbulent velocity is near the solar value (1km/s) throughout. 
Then the major parameters that characterize the spectra are $T_{\rm{eff}}$, {\logg} and $T_{\rm{cr}}$. 

\begin{figure}
  \begin{center}
  \epsscale{2}
  \plottwo{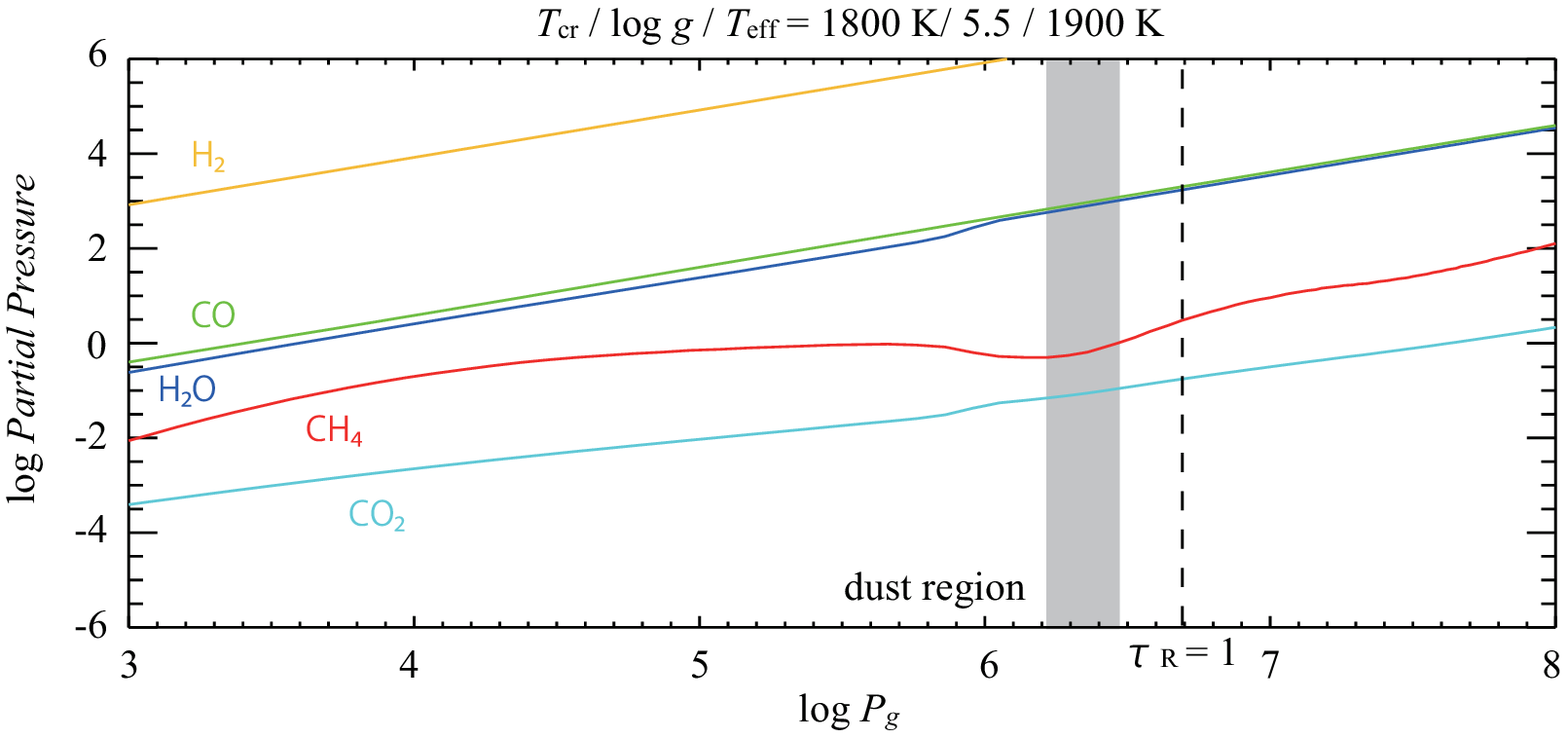}{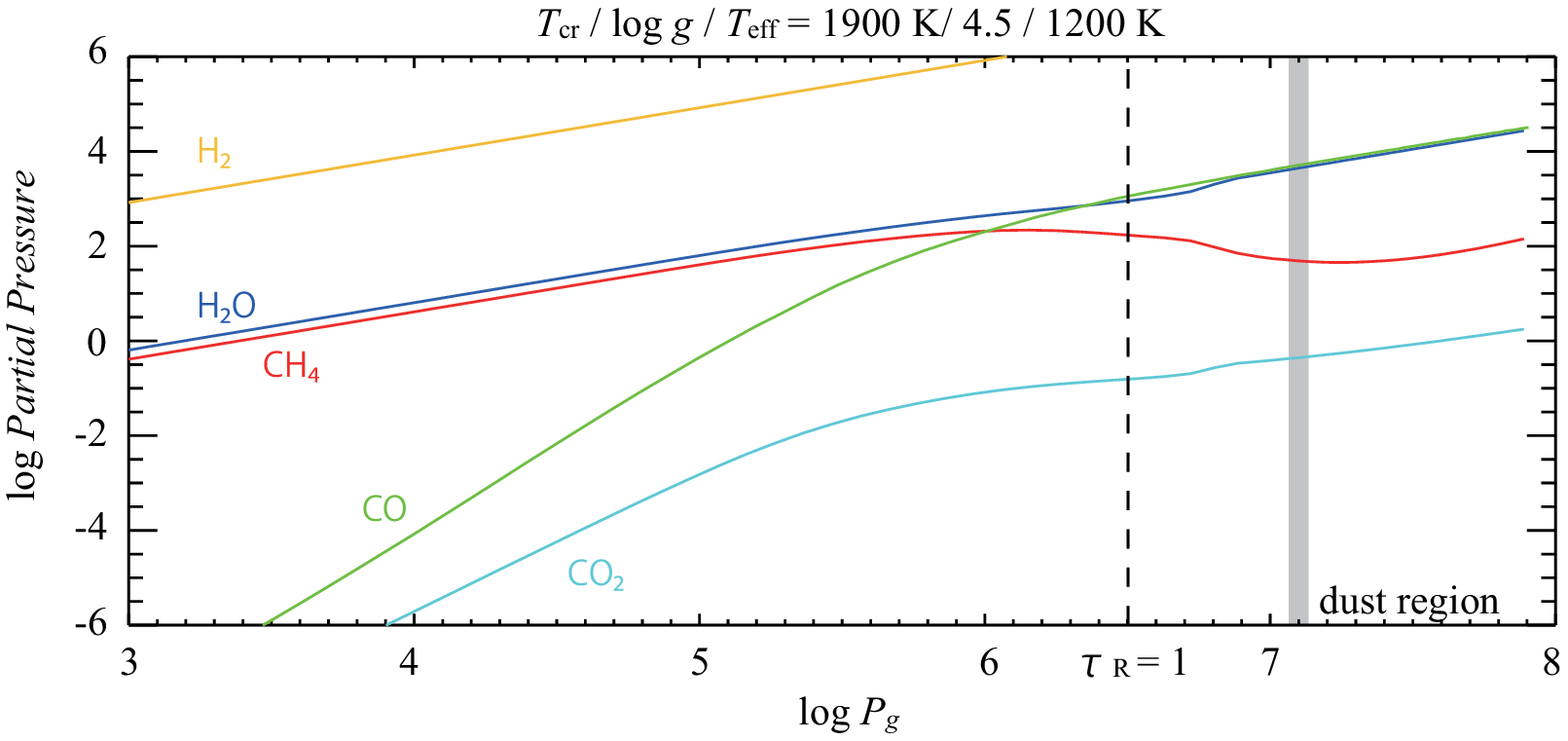}
  \end{center}
  \caption{Partial pressures of major molecules in the photospheres of brown dwarfs are plotted against total gas pressure.   
  The model  shown in the upper panel is with parameters of ({\Tcr}/{\logg}/{\Teff})=(1800~K/5.5/1900~K) corresponding to a L4 type.
  The lower panel is the model for ({\Tcr}/{\logg}/{\Teff})=(1900~K/4.5/1200~K) corresponding to a T5. 
  The dust layer is indicated in gray color. 
  We see that the CO abundance in T5 photosphere is rapidly decreasing toward the surface, log $P_g \sim 3.0$ dyn cm$^{-2}$, 
  and {\CHf} takes over the position of the most abundant carbon containing molecule.  
  } \label{ld}
\end{figure}

The two plots in Figures \ref{ld} show examples of the models corresponding to L and T dwarf with the parameters ({\Tcr}/{\logg}/{\Teff})=(1800~K/5.5/1900~K) and (1900~K/4.5/1200~K), respectively.  
These figures show profiles of molecular abundances along the position in the photosphere. 
The abundances change according to the difference of {\Teff}. 
We see that CO abundance in the photosphere of the T-type dwarf is significantly smaller than in the L-type photosphere, 
and instead {\CHf} becomes the major ingredient.  
We also show the dust layer ($T_{\rm{cr}}  < T <  T_{\rm{cond}}$) in Figures \ref{ld}. 
We see that the dust layer is located deeper in typical T dwarf photospheres than those in typical L dwarf photospheres. 
The dust in L dwarf photospheres exists in smaller Rosseland mean tau region than those in T dwarf photospheres, 
thus the dust affects an L-type spectrum more than a T-type spectrum. 

To fit the {\AKARI} spectra, we examine the cases of 700 $\le$ $T_{\rm{eff}}$ $\le$ 2200~K in 100~K grid, log $\textit{g}$=4.5, 5.0 and 5.5, $T_{\rm{cr}}$=1700, 1800, 1900~K and $T_{\rm{cond}}$ (no dust layer), 
i.e. a total of 16 $\times$ 3 $\times$ 4 = 192 cases.

\subsection{Fitting evaluation}
\label{fittingtechnique}
Our purpose of this analysis is to derive the most probable physical parameters for the {\AKARI} sample. 
The wavelength region of the {\AKARI} data gives us information about {\CHf}, CO and {\COt} molecules in the brown dwarf photospheres, 
and reflects the photospheric temperature relatively free from dust extinction. 
We use the {\AKARI} spectral data principally to derive the physical parameters via model fitting, 
but there are some technical hurdles to overcome, as discussed below.
Thus we use the shorter wavelength spectra (IRTF/SpeX and UKIRT/CGS4 data) to supplement our analysis. 
In this section we introduce the shorter wavelength spectra firstly. 
Then we explain the problem on model fitting and the fitting strategy to overcome the problem. 

\subsubsection{IRTF/SpeX data}
\label{spex}
The NASA Infrared Telescope Facility (IRTF) on Mauna Kea, Hawaii, is a 3.0 meter telescope optimized for infrared observations. 
SpeX is a medium-resolution spectrograph covering 0.8--5.4~$\mu$m on board  IRTF. 
The superiority of SpeX is the capability to provide maximum simultaneous wavelength coverage. 
A high throughput prism mode that uses single order long slit (60 arcsec) provides the spectral resolution $\lambda / \Delta \lambda = R \sim$ 100 for 0.8--2.5~$\mu$m. 
Using prism cross-dispersers (for 15 arcsec-long slits), $R$ becomes 1000--2000 across 0.8--2.4~$\mu$m, 2.0--4.1~$\mu$m, and 2.3--5.5~$\mu$m. 

Almost all brown dwarfs in our samples have been observed  by \citet{Burgasser_2004, Burgasser_2006a, Burgasser_2008, Burgasser_2010}, 
\citet{Burgasser_2007} and \citet{Cushing_phd} with SpeX. 
Nine data sets have been obtained using its low-resolution prism-dispersed mode with resolutions of 75--200 depending on the used slit-width. 
For these nine objects, we retrieve the data from \textit{The SpeX Prism Spectral Libraries} built by Adam Burgasser\footnote{URL; http://pono.ucsd.edu/$\sim$adam/browndwarfs/spexprism/ }. 
Only SDSS~J0539--0059 was unpublished\footnote{This data is now included in \textit{The SpeX Prism Spectral Libraries}.} and the data were obtained from Michael Cushing (private communication). 
Six other sources have been observed by SpeX using its short wavelength cross-dispersed mode with resolutions of 1200--2000, depending on the used slit-width. 
We obtained these data from the \textit{IRTF Spectral Library} maintained by Michael Cushing\footnote{URL; http://irtfweb.ifa.hawaii.edu/$\sim$spex/IRTF$\_$Spectral$\_$Library/}.

\subsubsection{UKIRT/CGS4 data} 
\label{ukirt}
SDSS~J1446+0024 has not been observed with SpeX. 
A spectrum in 1.0--2.5~$\mu$m was obtained with UKIRT/CGS4 (Geballe et al. 2002). 
CGS4 is a 1.0--5.0~$\mu$m multi-purpose grating spectrometer which was mounted on the 3.8 m United Kingdom Infrared Telescope (UKIRT), 
which is sited on Mauna Kea, Hawaii. 
CGS4 has four gratings. 
The data of SDSS~J1446+0024 were obtained using 40 line/mm grating that provided the resolution of 300--2000, 
which are defined by 400 $\times$ $\lambda$. 
The wavelength coverage of this observation is 1.03--1.34~$\mu$m and 1.43--2.53~$\mu$m. 
We obtained the spectral data of SDSS~J1446+0024 from Dagny Looper (private communication).

\subsubsection{Absolute Flux Calibration of Short Wavelength Spectral Data}
\label{calspex}
Since nine SpeX data of {\AKARI} samples are normalized at 1.25~$\mu$m, 
we calibrate their absolute fluxes using the $J$ band photometric data (hereafter $J_{\mathrm{p}}$) given by \citet{Leggett_2002a} and Knapp et al. (2004) listed in Table~\ref{maggol_l}. 
These $J_{\mathrm{p}}$ were taken with the MKO filter. 
SIMP J0136+0933 was not observed with the MKO filter,
and we use 2MASS~$J$ band photometric data (\citealt{Cutri_2003}) for the flux calibration of this object. 
The 2MASS~$J$ band magnitude is also listed in 8th-column of Table~\ref{maggol_l}.
We estimate the $J$ band flux from SpeX data ($J_{\mathrm{SpeX}}$) with the MKO or 2MASS~$J$ band filter transmission function $T$ by calculating equation (\ref{mkofilter}), 
\begin{equation}
\label{mkofilter}
J_{\mathrm{SpeX}}= \frac{\sum F_{\nu} T(\nu) \Delta \nu}{\sum T(\nu) \Delta \nu}. 
\end{equation}
After that, we scale the absolute flux of SpeX spectral data with the ratio of $J_{\mathrm{p}}/J_{\mathrm{SpeX}}$. 

\begin{figure}
\epsscale{.80} 
\begin{center}
   \plotone{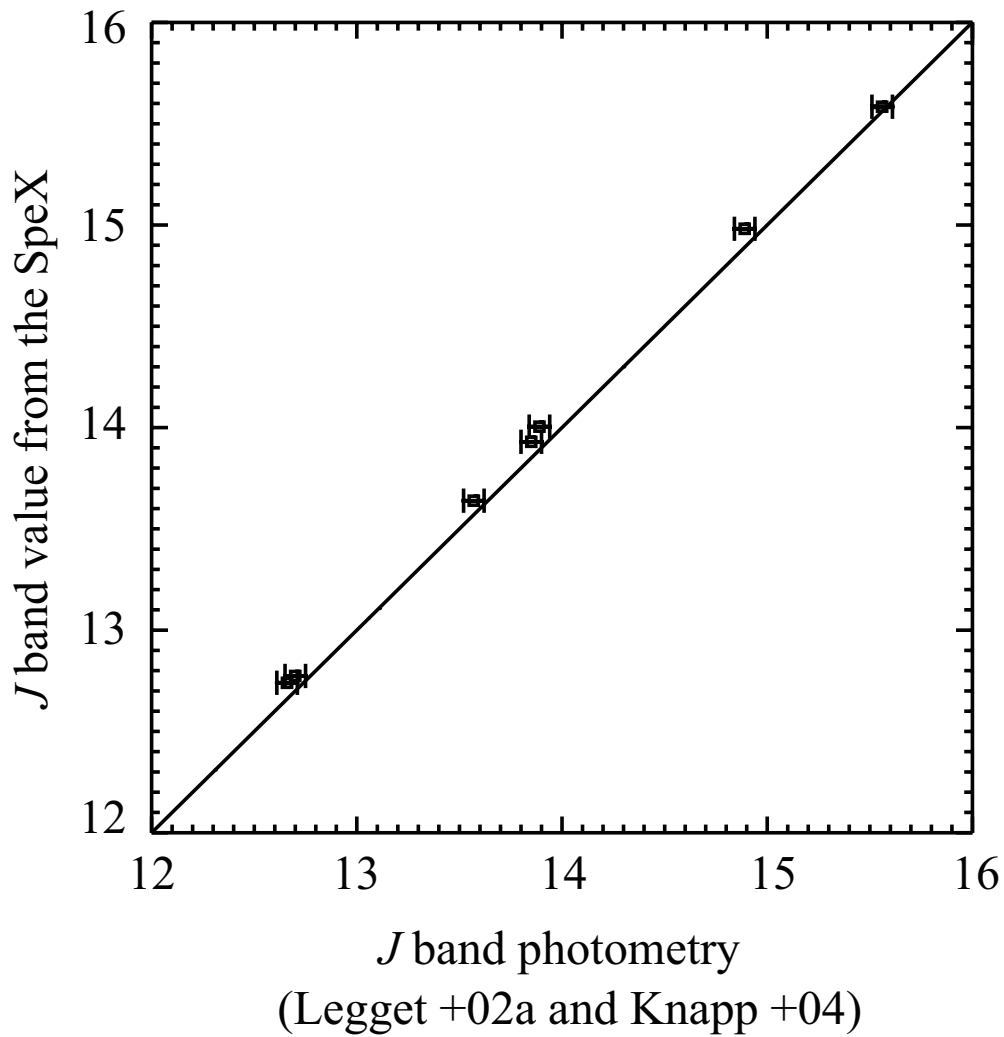}
\end{center}
\caption{
The $J$ band fluxes measured on the IRTF/SpeX and UKIRT/CGS4 spectra
are compared with the photometry data by Leggett et al. (2002a) 
and Knapp et al. (2004).
Six SpeX spectra are calibrated by Cushing et al. (2004),
and one SGC4 spectrum is by Geballe et al. (2002), respectively.
Two measurements of these 7 objects are consistent to within 10~\%.
} 
\label{jgol_jspex}
\end{figure}

The absolute flux levels of six objects observed by SpeX and one object observed by the UKIRT/CGS4 were calibrated with the 2MASS~$J$ band photometry data by \citet{Cushing_2004} and \citet{Geballe_2002}, respectively. 
We derive the $J$ band flux from the calibrated spectra to compare with $J_{\mathrm{p}}$ from \citet{Leggett_2002a} and \citet{Knapp_2004}.
They are confirmed to be consistent within 10$\%$.  
Figure~\ref{jgol_jspex} shows the results. 

\begin{figure}
\epsscale{.80} 
\begin{center}
   \plotone{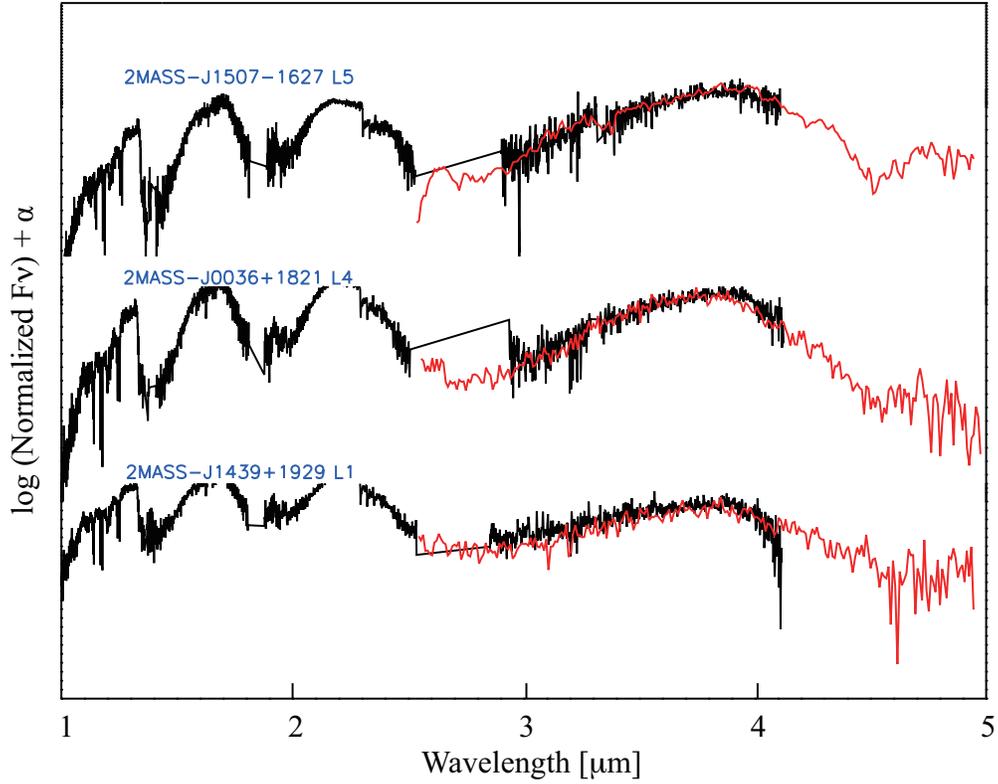}
\end{center}
\caption{The comparison of SpeX spectral data (black) and {\AKARI} spectra (red) for 2MASS~J1439+1929, 2MASS~J0036+1821 and 2MASS~J1507--1627. 
Both {\AKARI} and SpeX/CGS4 data of each object are calibrated independently and scaled by the same factor on the plot.  
} 
\label{spexspectra2}
\end{figure}

The spectra of three objects,  2MASS~J1439+1929, 2MASS~J0036+1821 and 2MASS~J1507--1627, were observed by SpeX in 0.81--4.11~$\mu$m simultaneously with a gap over 2.53--2.85~$\mu$m. 
They are calibrated by \citet{Cushing_2004}. 
In Figure~\ref{spexspectra2} their {\AKARI} spectra and SpeX spectra are plotted in red and black line, respectively. 
Both {\AKARI} and SpeX data of each object are calibrated independently and scaled by the same factor on the plot.  
We find that two spectra are consistent within 10 $\%$, which is the uncertainty level of the {\AKARI} spectra. 

\subsubsection{Step1 -- Fitting the {\AKARI} Spectra}
\label{firstfit}
Our model fitting consists of two processes. 
As the first step we compare the models with observed {\AKARI} spectra between 2.5 and 4.15~$\mu$m. 
We use the data only over 2.5--4.15~$\mu$m, not up to 5.0~$\mu$m for fitting, 
because we know that the current model does not explain the observed spectra beyond 4.15~$\mu$m where the absorption bands of {\COt} at 4.2~$\mu$m and CO at 4.6~$\mu$m are present (\citealt{Yamamura_2010}, \citealt{Tsuji_2011}). 
We follow \citet{Cushing_2008} and evaluate the goodness of the model fitting by the statistic G$_{k}$, defined as 

\begin{equation}
\label{gk}
G_{k} = \frac{1}{n-m}\sum_{i=1}^n \omega_{i}  \left( \frac{f_{i} - C_{k}F_{k,i}}{\sigma_{i}} \right)^2,
\end{equation}
where $n$ is the number of data points; $m$ is degree of freedom (this case $m=3$); $\omega_{i}$ is the weight for the $i$-th wavelength points 
(we give the weight as $\omega_{i}$ = 1 for all data points because of no bias within each observed spectrum); 
$f_{i}$ and $F_{k, i}$ are the flux densities of the observed data and $k$-th model, respectively; 
$\sigma_{i}$ are the errors in the observed flux densities 
and $C_{k}$ is the scaling factor given by 

\begin{equation}
\label{scalingfactor}
C_{k} = \frac{\sum \omega_{i} f_{i} F_{k,i}/\sigma_{i}^2}{\sum \omega_{i} {F_{k,i}}^2/{\sigma_{i}}^2}.
\end{equation}
$G_{k}$ is equivalent to reduced $\chi ^{2}$, since we adopt $\omega_{i}$ = 1 in our analysis. 

\begin{figure}
\begin{center}
   \plotone{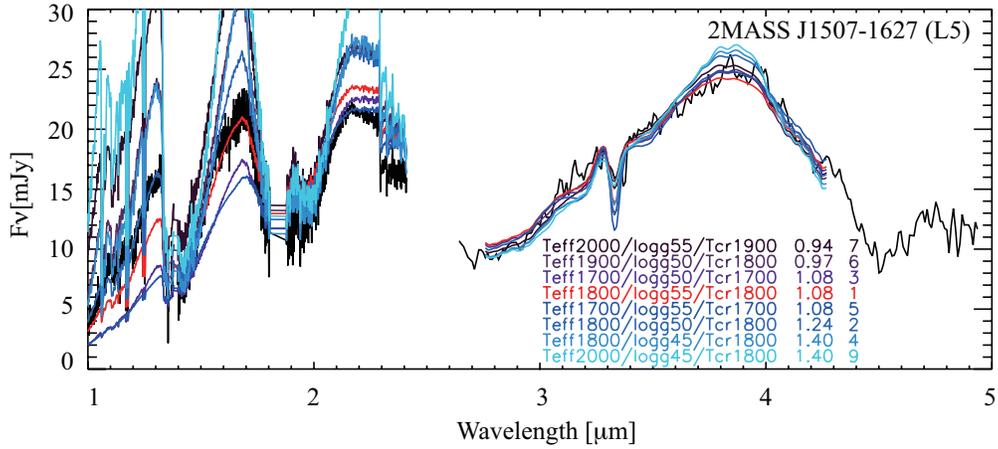}
\end{center}
\caption{The spectra of 2MASS~J1507--1627(L5) and models. 
The degeneracy in the fitting of {\AKARI} spectrum is solved by adding the SpeX data of the shorter wavelength range. 
The model spectrum in red color is the best fit model for this object. 
The order of goodness of the fit to the SpeX data are shown on the right of the legend.} 
\label{degeneracy2}
\end{figure}

It is difficult to determine a unique best fit model for each {\AKARI} object because of the large error associated with the {\AKARI} spectral data.
In general, when the reduced $\chi ^2$ (=$G_{k}$) is 1--2, the model is regarded to fit the observed data well. 
However, in our case $G_{k}$ easily falls below unity and we have too many $``$good fit$"$ models. 
This degeneracy is demonstrated in Figure~\ref{degeneracy2} for the {\AKARI} spectra of 2MASS~J1507--1627 (L5). 
We see that many models have small ($\le$ 1.04) $G_{k}$ between 1700 to 2000~K of {\Teff}. 
The minimum $G_{k}$ is 0.94, and the second minimum is 0.97. 
The differences of $G_{k}$ between the models near the minimum are too small to determine the best fit model.

Therefore at this step, we select candidates of the best model with the following condition.  
\begin{equation}
\label{gplus1}
G_{min} \le G_{k} < G_{min} + 1, 
\end{equation} 
where $G_{min}$ is  the minimum $G_{k}$ value. 
$G_{min}$ is different for every object and is not always near unity, 
due to the difference in the error of the {\AKARI} observed spectra. 
We apply $G_{min}$ + 1 as an upper limit.
This criterion reasonably selects  5 $\sim$ 20 model candidates for almost all the observed data.  

\subsubsection{Step2 -- Constraining Models with the Short Wavelength Data}
To constrain the models from the candidates, we additionally use the spectra in the shorter wavelength range (1.0--2.5~$\mu$m) taken by IRTF/SpeX  and UKIRT/CGS4.  
It is only possible to constrain the model parameter uniquely
with the help of the short wavelength range spectra along with the {\AKARI} data. 
The wavelength region of the {\AKARI} data gives us information about molecules (=gas) in brown dwarf photospheres. 
On the other hand, data covering shorter wavelengths are the most sensitive to the presence of dust.

We calculate $G_{k}$ for the IRTF/SpeX or UKIRT/CGS4 (hereafter SpeX/CGS4) spectral data for the candidate models derived in Step~1. 
Since we have validated the absolute flux of the spectra to be accurate to better than 10 $\%$ (see section \ref{aflux}), 
we apply $C_k$ values derived in Step~1 (from the {\AKARI} data) to Step~2. 
Figure \ref{degeneracy2} shows that the degeneracy appeared in the {\AKARI} wavelengths is resolved in the shorter wavelengths. 
Results of the model fitting through these processes are shown in Table~\ref{fittable}.

We do not fit the {\AKARI} data and the SpeX/CGS4 data simultaneously. 
This is because the error of two data sets are very different. 
The average relative error of the {\AKARI} spectra is about 10 $\%$ (Ohyama et al. 2007), 
while that of the SpeX and CGS4 data is below 0.05 $\%$ (Reyner et al. 2009). 
This difference would give much higher weight to the SpeX/CGS4 data in the fitting evaluation in Equation (\ref{gk}). 
Actually, while the reduced $\chi ^2$ (= $G_{k}$) values of the {\AKARI} data are between 0.1 and 100, 
those of the SpeX/CGS4 data are between 100 and 5000.
Therefore, we decided to use the {\AKARI} spectra first and use the SpeX and CGS4 spectra in the second step.

\subsection{Uncertainty of the Model Fitting}
\label{uncertainty}
Here we discuss the uncertainty of the best fit model parameters. 
The models we use for the current analysis are calculated on the 100~K (for {\Teff} and {\Tcr}) and 0.5 dex (for {\logg}) grid, 
and the uncertainty should be no better than the grid spacing. 
To estimate the uncertainty we change one of {\Teff}, {\logg} and {\Tcr} by one grid from the best fit value, 
and search for the $``$restricted best$"$ model by changing other two parameters following the same manner through Step~1 and 2. 
If we do not find any models satisfying $G_{min} \le G_{k} < G_{min} + 1$ (here $G_{min}$ is taken from the all parameter space in Step~1), 
the uncertainty of the parameter should be smaller than the grid spacing. 
When the best parameter is already on the edge of the parameter space; i.e., {\Teff}  = 700 or 2200~K, {\logg} = 4.5 or 5.5 and {\Tcr} = 1700~K or {\Tcond}, 
we only run the test on the available side of the parameter grid. 

\begin{figure}
\begin{center}
   \plotone{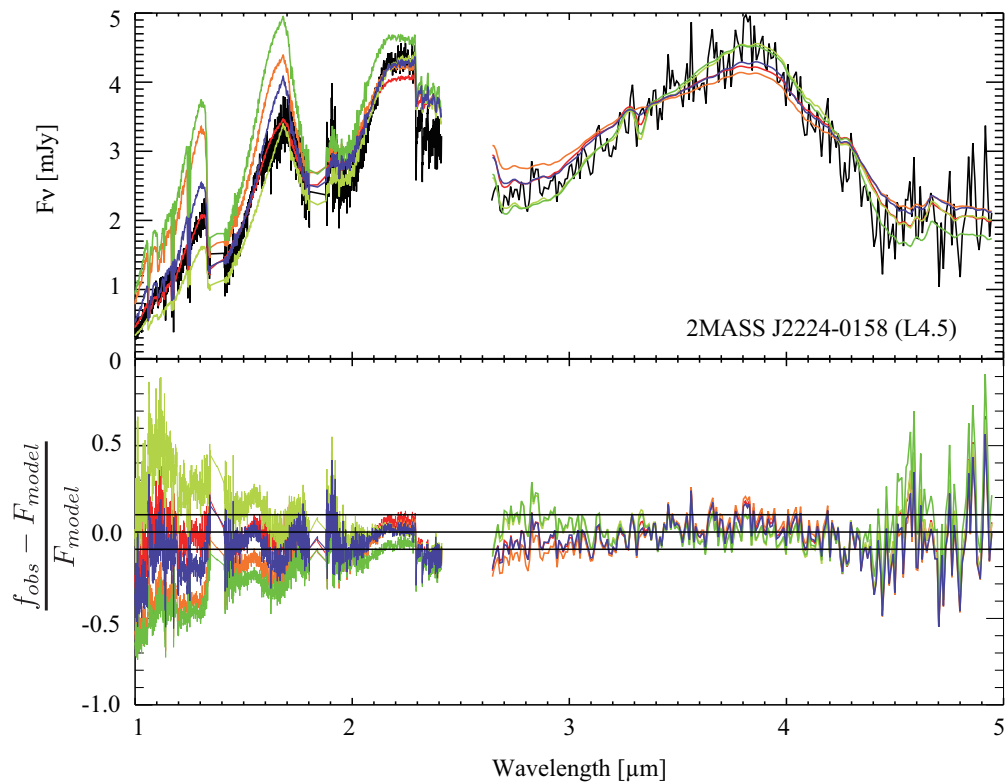}
\end{center}
\caption{The comparison of the $``$restricted best$"$ models with observed spectrum for 2MASS~J2224--0158 (L4.5). 
The real best fit model is in red and the $``$restricted best$"$ models for {\Teff} (+100~K and --100~K), {\Tcr} (+100~K and --100~K) and {\logg} (+0.5 and --0.5) are in orange, yellow, green, light blue, blue, and purple, respectively. 
We see that the there are large differences in the $J$ and $H$ bands. 
} 
\label{examplec}
\end{figure}

We show a detailed example for 2MASS~J2224--0158 (L4.5), whose best model is ({\Tcr}/{\logg}/{\Teff})=(1700~K/5.5/1800~K), in Figure~\ref{examplec}.  
We see large differences in $J$ and $H$ bands in some $``$restricted best$"$ model spectra. 
We derive a factor to further scale the $``$restricted best$"$ model from that given by $C_k$ to adjust the observed spectra in the $J$ and $H$ band region (1.01--1.81~$\mu$m) using Equation \ref{scalingfactor} (hereafter $C_{J,H}$). 
We exclude the $``$restricted best$"$ models if the factor, $C_{J,H}$, is more than 1.10 or less than 0.90, regarding the uncertainty of the SpeX/CGS4 absolute flux.
For example $C_{J,H}$ for the $``$restricted best$"$ models for {\Teff} ($+100$~K and $-100$~K), {\Tcr} ($+100$~K) and {\logg} ($-0.5$) of 2MASS~J2224--0158 (L4.5) are 0.74, 1.12, 0.65, 0.88, respectively. 
Thus any $``$restricted best$"$ model is invalid for 2MASS~J2224--0158 (L4.5), i.e., the uncertainty is smaller than one grid. 

We find that the $``$restricted best$"$ models for the case of changing {\Teff} by $\pm$100~K and {\Tcr} by +100~K exhibit a noticeable change from the real best models.  
On the other hand, the case of changing {\Tcr} by --100~K and {\logg} result only in minor differences. 
We further continue the test of changing each parameter by two grids from the real best fit model. 
Almost all the $``$restricted best$"$ models do not stay between 0.90 and 1.10 any longer, except for the case of changing two grids of {\logg}. 
We conclude that our best fit model parameters are determined as good as one grid, and at worst two grids of each parameter for {\Teff} and {\Tcr}, 
but $C_{J,H}$ for the case of changing two grids of {\logg} still stays between 0.90 and 1.10. 
However, we can not change the grid of {\logg} any more. 
Therefore the uncertainty of {\logg} is not determined well for some objects. 
The uncertainties for each object are listed in superscript and subscript in Table~\ref{fittable}.

\section{Brown Dwarf Atmospheres along with the Spectral Types}
\label{resultucm}

\subsection{Comparison of the Observed and the Best Fit Model Spectra}
\label{resultcomucm}
\begin{figure}
      \plottwo{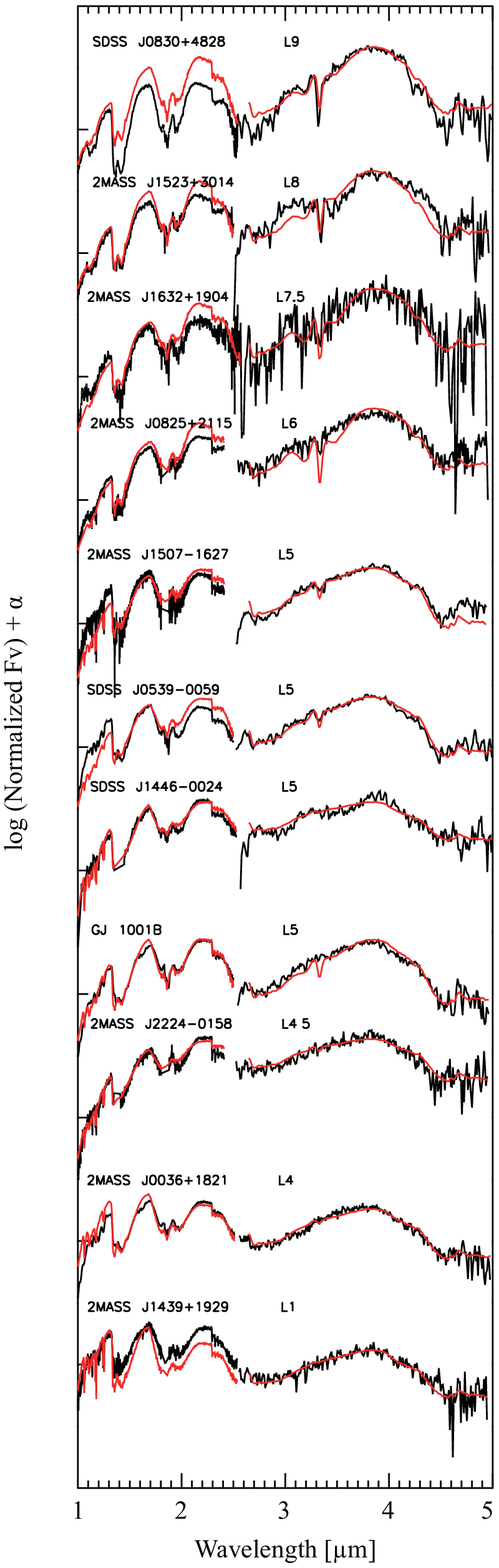}{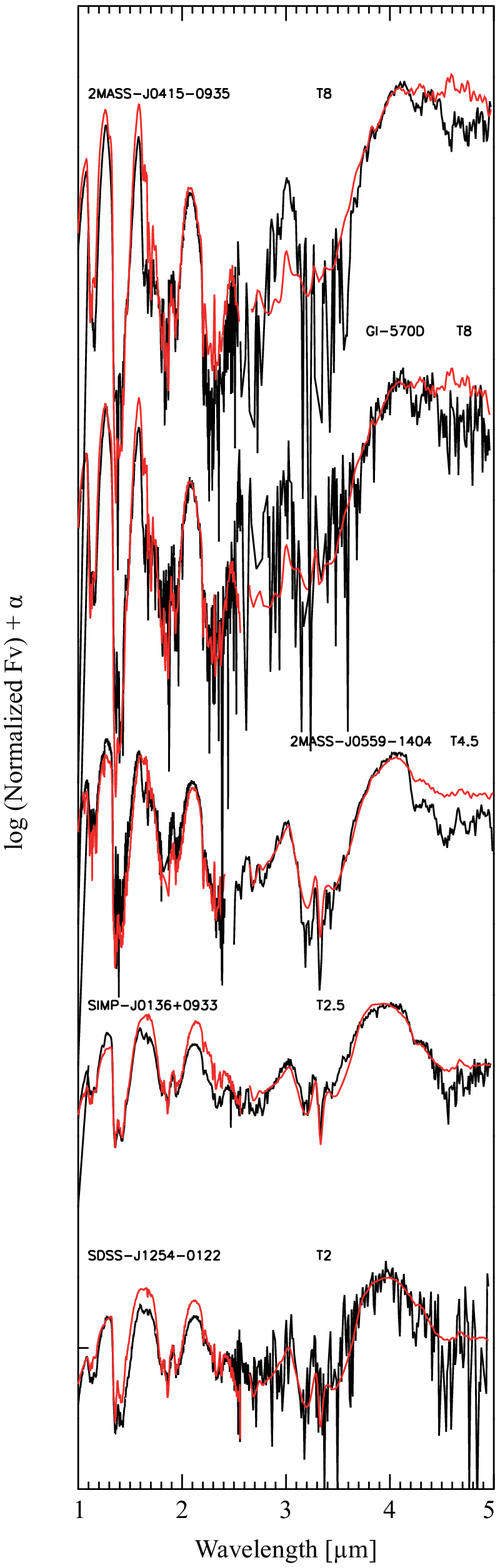}
\caption{The fitting result. 
The black lines are the observed data and red lines are the best fit model spectra. 
The spectra between 2.5 and 5.0~$\mu$m are taken by {\AKARI} and those between 1.0 and 2.5~$\mu$m are taken by IRTF/SpeX (except for SDSS~J1446+0024 observed by UKIRT/CGS4). } 
\label{fit}
\end{figure}

The spectra of observed and best fit models are compared in Figure~\ref{fit}. 
The model spectra generally explain the observed spectral features well, except for some objects noted below. 
We see that the fit in the entire {\AKARI} region is fairly good, except for the wavelength region longer than 4.15~$\mu$m in mid- to late-T dwarfs, which we do not take into account in the fitting evaluation. 
We were aware that CO and {\COt} bands are not reproduced by the current UCM. 
Actually, the model fit including this wavelength range makes the fit of the overall {\AKARI} range even worse.

A noticeable deviation is seen in the late-T dwarfs around 3.0~$\mu$m where {\HtO} and {\CHf} absorption features overlap. 
The flux density around the 3.0~$\mu$m region in the model spectra of three late-T dwarfs is too low in comparison to that in the observed spectra. 
The {\CHf} absorption at 1.6~$\mu$m in the model spectra of these late-T dwarfs is always significantly weaker than that in the observed spectra. 
For other brown dwarfs model spectra sometimes do not explain  the strength of {\CHf} $\nu_3$ absorption band around 3.3~$\mu$m. 
We also find that {\COt} absorption band in the model of late-L to T dwarfs is sometimes too deep and sometimes too shallow compared to the observations. 
The fit in the SpeX/CGS4 region is good in an overall sense, except for five late-L and two early-T dwarfs. 
The $H$ and $K$ band flux density in the model spectra of these five dwarfs are higher than that in the observed spectra.

The high flux level around the 3.0~$\mu$m region in the observed spectra indicates that actual photospheric temperature is higher than that of the models. 
On the other hand, the result of stronger {\CHf} absorption band at 1.6~$\mu$m in the observed spectra of late-T dwarfs indicates that the temperature in the photosphere of these dwarfs should be lower than observed. 
This contradiction implies that the thermal structure of the objects derived by the model is still not perfect.  
The deviation of the 3.3~$\mu$m {\CHf} for mid- to late-L dwarfs may also indicate the incompleteness of thermal structure in the model.  
We have to consider the mechanism to improve the thermal structure to reproduce the observation data. 
The second possible reason of discrepancy between the model and observations is incompleteness of the line lists.
It is known that the line lists of polyatomic molecules, such as {\HtO}, {\CHf}, or {\COt}, are still not perfect. 
The effects of line list should be investigated, however it is not likely that these discrepancies are caused only by the incomplete line lists.
A third option would be the elemental abundance. 
The discrepancy of {\COt} absorption in the observed spectra may indicate that the elemental abundances in the photospheres of these objects are different from the assumptions in the model.  
Excess of {\COt} absorption of SDSS~J0830$+$4828~(L9), 2MASS~J0559--1404~(T4.5) and 2MASS~J0415--0935~(T8) were discussed by Tsuji et al. (2011) in terms of enhanced C and O elemental abundances. 
Recently, \citet{Madhusudhan_2011} reported an anomaly in the {\HtO} 
and {\CHf} abundances compared to the solar abundance chemical 
equilibrium model prediction in the atmosphere of the hot-Jupiter WASP-12b.
They suggested that the abundance of these molecules can be explained
if the carbon-to-oxygen ratio [C]/[O] in this planet's atmosphere is
much greater than the solar value ([C]/[O]=0.54), i.e., [C]/[O]$>1$.
Although the structure of atmosphere may be different in planets
and brown dwarfs, these results are consistent with our conclusion;
the elemental abundance is an essential parameter of brown dwarf
/ planet atmosphere and should be carefully considered in future
studies.
We extend the study of possible elemental abundance variations among brown dwarfs using model atmospheres and the {\AKARI} data in a forthcoming paper (Sorahana et al. in prep).

The $H$ and $K$ band flux densities in the model spectra of late-L and early-T dwarfs are always higher than those in the observed spectra. 
Since the wavelength range of SpeX/CGS4 is the most sensitive to the dust extinction, 
we can evaluate the dust amount from the spectra of this wavelength region. 
The effect of dust extinction turned out to be small in the late-L to early-T dwarfs.
Less warming up effect by the dust is expected. 
This argument indicates that an increase of the dust and the inner temperature are overestimated in the models as compared to actual photospheres. 
Since dust opacity relies on the composition, grain size distribution and amount, we shall confirm the effects of these quantities in the UCM. 
We also propose that a self-consistent, more realistic theory of condensation and sedimentation in the atmospheres is essential in future brown dwarf atmosphere models.

\subsection{Model Parameters and Spectral Type}
\begin{figure}
\epsscale{.40}
\begin{center}
    \plotone{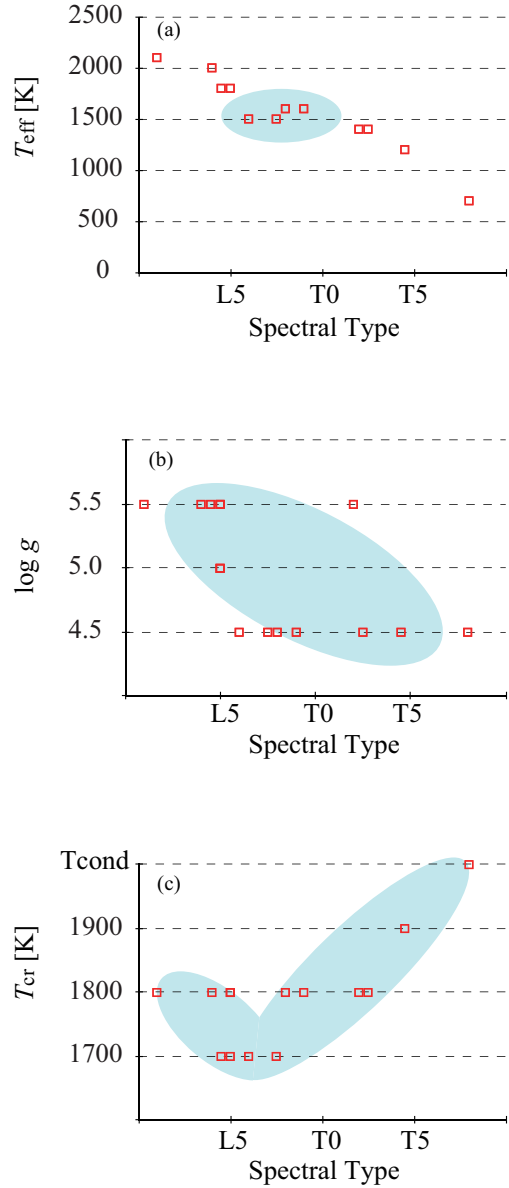}
\end{center}
\caption{The best fit model parameters versus spectral type, (a) {\Teff}, (b) {\logg}, and (c) {\Tcr}. 
Objects with the same spectral type may overlap when they have the same parameters. 
We see that spectral type is in the sequence of {\Teff} except for the late-L dwarfs. {\logg} decreases toward the late spectral types. 
{\Tcr} is minimum for mid- to late-L types. } 
\label{aspara}
\end{figure}

Parameters of the best fit models are shown in Table~\ref{fittable}, 
and the parameters are plotted in Figure~\ref{aspara} with respect to the spectral types. 
We see that the spectral types are in principle in the sequence of {\Teff} (Figure~\ref{aspara} (a)). 
{\Teff} in late-L dwarfs have approximately the same value. 
{\logg} shown in Figure~\ref{aspara} (b) appears to be associated with the spectral types, 
but with large uncertainty as shown in section \ref{uncertainty}. 
For {\Tcr} shown in Figure~\ref{aspara} (c), we see a decreasing trend from L1 to L6, then {\Tcr} changes and increasing for the objects later than L6.
The uncertainties of {\Tcr} of 2MASS~J1523+3014 (L8) and SDSS~J0830+4828 (L9) are relatively large. 
For the spectra near the L/T transition it is very difficult to fit the spectra over a wide wavelength range with a model spectrum. 
Even the best fit model spectrum deviates from observed spectrum in many points. 
However, the spectral shape of the best fit model agrees with the observed one much better than any $``$restricted best$"$ models by eye, and we are convinced that the best fit model parameters are secure, 
i.e., {\Tcr} has a minimum at L6. 

The result of almost constant {\Teff} in late-L dwarfs has already been pointed out by past studies. 
Tsuji and Nakajima (2003) argued that little change in {\Teff} would be caused by cloud migration from optically thin upper regions to the thick inner regions. 
We confirm this hypothesis from the derived {\Tcr} in Figure~\ref{aspara} (c). 
From our fitting analysis the dust effect appears to become larger from early- to mid-L dwarfs. 
This result is consistent with the $J - K$ color shown in Figure~\ref{jkcolor}. 
Mid-L dwarfs should have the largest amount of dust.  
The trend of {\Tcr} along their spectral types shows that the contribution of dust to the spectra becomes smaller from late-L dwarfs to late-T dwarfs. 
Our result attests to the suggestion in \citet{Tsuji_2003}. 

\begin{deluxetable}{lclll}
\tabletypesize{\scriptsize}
\tablecaption{Best fit model parameters and their uncertainty derived by fitting the {\AKARI} and the SpeX/CGS4 data. \label{fittable}}
\tablewidth{0pt}
\tablehead{
\colhead{Object Name} &\colhead{Sp.Type} & \colhead{{\Tcr}[K]} &  \colhead{{\logg}}&  \colhead{{\Teff}[K]}
}
\startdata
2MASS~J1439+1929 &L1&1800$_{-100}$ &5.5$_{-1.0}$ &2100 \\
2MASS~J0036+1821 &L4& 1800 & 5.5  &2000$_{-100}$   \\
2MASS~J2224--0158 &L4.5& 1700 &  5.5 & 1800  \\
GJ~1001B &L5& 1800$_{-100}$ &  5.0$_{-0.5}$ & 1800  \\
SDSS~J1446+0024 &L5& 1700$^{+100}$   &  5.0$^{+0.5}$   & 1800   \\ 
SDSS~J0539--0059 &L5& 1800  & 5.5  &  1800  \\
2MASS~J1507--1627&L5&1800  &  5.5$_{-0.5}$ & 1800  \\
2MASS~J0825+2115 &L6&  1700& 4.5  & 1500  \\
2MASS~J1632+1904 &L7.5&1700  & 4.5$^{+0.5}$  &1500$^{+100}$   \\
2MASS~J1523+3014 &L8& 1800$_{-100}$ & 4.5$^{+1.0}$  &1600${^{+100}_{-100}}$   \\
SDSS~J0830+4828 &L9&  1800$_{-100}$ & 4.5$^{+0.5}$  & 1600$_{-100}$  \\
SDSS~J1254--0122 &T2& 1800  & 5.5  & 1400  \\
SIMP J0136+0933 &T2.5&  1800$_{-100}$& 4.5$^{+1.0}$  & 1400$^{+100}$ \\
2MASS~J0559--1404 &T4.5& 1900$^{+100}$ & 4.5$^{+0.5}$ &  1200$^{+100}$ \\
Gl~570D &T8& 2000 &  4.5 & 700  \\
2MASS~J0415--0935 &T8& 2000 &  4.5 & 700 
\enddata
\tablecomments{
The upper and lower uncertainties for each object are shown in the superscript and subscript, respectively.}
\end{deluxetable}

\subsection{Advantage of {\AKARI} Spectra to the Model Fitting}
In previous studies, physical parameters of brown dwarfs have been derived mainly from near-infrared (1.0--2.5~$\mu$m) spectra.
\citet{Tsuji_2004} attempted to interpret the near-infrared spectra (0.882--1.400~$\mu$m, 1.056--1.816~$\mu$m, and 1.850--2.512~$\mu$m) obtained with Subaru Telescope with the UCM. 
They reported that the overall SEDs and the strengths of the major spectral features are reasonably, but not perfectly, reproduced by the UCM.
In their fitting, dust extinction effects (most effectively 
contributing to $J$ and $H$ band) and molecular band
strengths were the key features in determining the physical
parameters from model candidates, 
but it is hardly realized that these features are all consistent with an observation. 
They commented that it was difficult to derive reliable physical parameters for each object by the UCM even though their fitting to the near-infrared spectra (only) can constrain the physical parameters in a certain range.

The spectra in 2.5--5.0~$\mu$m obtained by {\AKARI} should provide additional information on brown dwarf atmospheres, as the fundamental bands of major molecules, {\HtO}, CO, {\COt} and {\CHf} are located in this region with less effects of dust opacity.
These molecular bands may sample different part of the photosphere. 
By including the {\AKARI} spectra into the model analysis, we expect to improve the model fitting to reflect more global characteristics of the objects.
\citet{Cushing_2008} also mentioned that the model fitting to the broader wavelength range results in more reliable physical parameters, even though the fits to the narrower wavelength region can be better.

\begin{figure}
\epsscale{.70}
\begin{center}
   \plotone{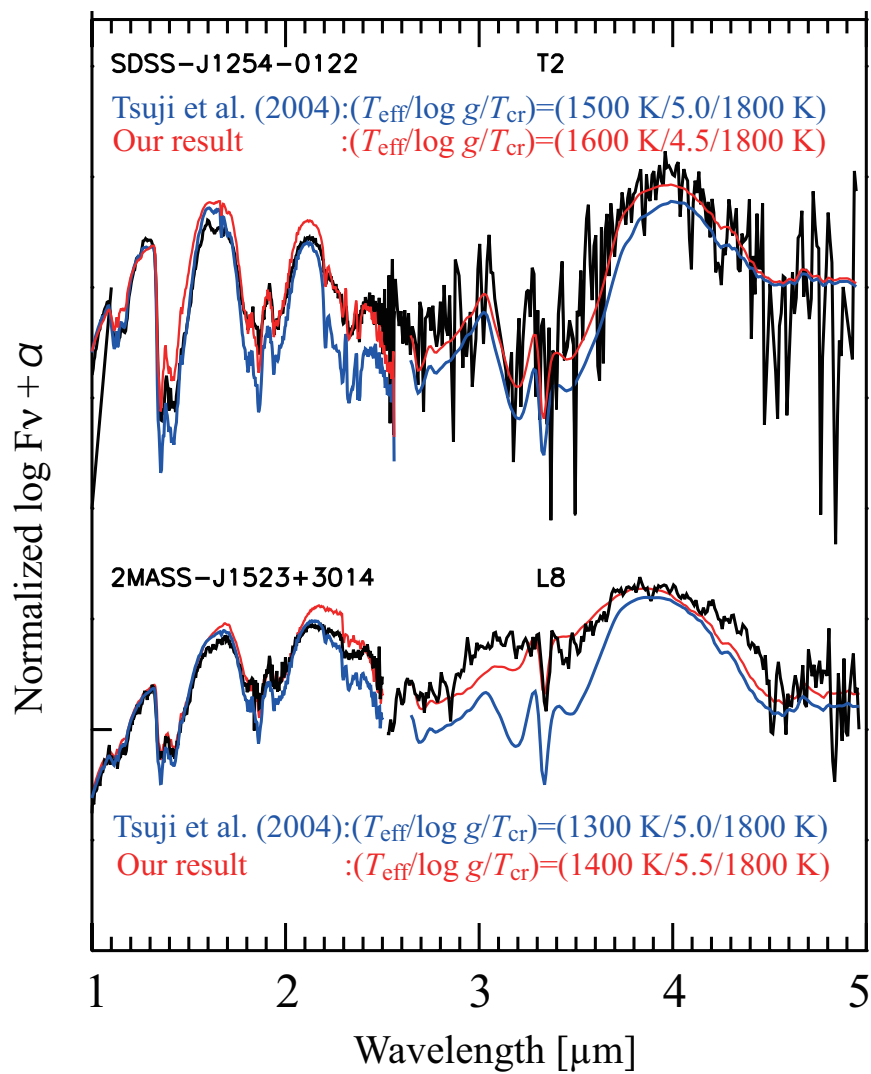}
\end{center}
\caption{The best fit model spectrum in the current study (red) and 
that from Tsuji et al. (2004) (blue) are compared with the 
AKARI+SpeX observed spectra (black) of two brown dwarfs, 
2MASS J1523+3014 and SDSS J1254--0122.
It is seen that the near-infrared part ($\leq 2.5$~$\mu$m) is reasonably well reproduced by both the models, but {\AKARI} spectra at longer wavelength is much better explained by the current model, showing that the {\AKARI} spectra give further constraint to the model parameters.
 } 
\label{comspectsuji}
\end{figure}

To demonstrate the benefit of the {\AKARI} spectra, we compare the observed spectra of two brown dwarfs, 2MASS~J1523$+$3014 (L8)and SDSS~J1254$--$0122 (T2), with the models derived by the current study and by \citet{Tsuji_2004} 
in Figure \ref{comspectsuji}.
We see that both model spectra reasonably reproduce the near-infrared part of the observed spectra. However, the {\AKARI} spectra can be explained only with our models.

This result indicates that the parameters derived only from near-infrared spectra might have large uncertainties, and the broader wavelength data including {\AKARI} spectra are
essential to understand the nature of brown dwarf atmospheres.
We find that {\Teff} derived by our model fitting is generally $\sim$100 K higher than that of \citet{Tsuji_2004}  for the same spectral type objects.
It is probably because of the fact that molecular bands in the {\AKARI} wavelength range help to distinguish between the effects of {\Teff} and {\Tcr}.

\section{Conclusion}
We study the presence of 4.6~$\mu$m CO,  4.2~$\mu$m CO$_{2}$ and 3.3~$\mu$m CH$_{4}$ bands in 16 {\AKARI} brown dwarf spectra over the wide range of spectral types. 
We confirm that the CH$_{4}$ band appears in the sources as early as L5, 
and find that the band is seen in only two of four L5 dwarfs in our sample. 
Their $J-K$ color indicates that the appearance of {\CHf} band in two L5 dwarfs is related to the dust abundance. 
The CO 4.6~$\mu$m band appears in the spectra of all spectral types until late-T dwarfs. 
The fact that CO generally exists in all brown dwarf atmospheres is very important,
since it is confirmed that the deviation of molecular abundance in brown dwarf atmospheres from the LTE prediction is a common feature.
We need further consideration on this problem in future works. 
We detect $\mathrm{CO_2}$ absorption band at 4.2~$\mu$m in the spectra of late-L and T type dwarfs.
These detections indicate that the $\mathrm{CO_2}$ molecule is generally in the atmospheres of these dwarfs.

We analyze the {\textit{AKARI}} spectra using the UCM. 
We derive the three physical parameters, effective temperature {\Teff}, critical temperature {\Tcr}, and surface gravity {\logg}, of 16 sources by systematic model fitting. 
We investigate how the spectral type correlate with the parameters. 
We find that the spectral type follows a sequence of {\Teff}, except for the late-L dwarfs, 
for which the spectral type is a sequence of {\Tcr}, the parameter related to the effects of dust. 
This result confirms expectations from past studies. 
{\AKARI} give us these new insights on brown dwarf atmospheres in the new spectral range, 2.5--5.0~$\mu$m. 
We also find important problems, which are not explained with the current brown dwarf atmosphere model. 
Therefore, the models need improvements and the {\AKARI} spectra should be analyzed in further detail. 

We thank to the anonymous referee for critical reading of our manuscript
and constructive suggestions. We are
grateful to Dr. Poshak Gandhi for his careful checking our
manuscript and suggestions to improve the text.
This research is based on observations with {\AKARI}, a JAXA project with the participation of ESA. 
We thank Prof. Takashi Tsuji for his kind permission to access the UCM and for helpful suggestions. 
Dr. Adam Burgasser, Dr. Michael. Cushing, and Dr. Dagny. Looper generously provided me (Sorahana) observed data, along with warm encouragement.
We acknowledge JSPS (PI: S. Sorahana) and JSPS/KAKENHI(c) No. 22540260 (PI: I. Yamamura).

\bibliography{sora1012}

\end{document}